\documentclass[aps,amsmath,amssymb,twocolumn,showpacs,pra]{revtex4-1}
\usepackage{graphicx}
\usepackage{bm}

\allowdisplaybreaks
\input{Qcircuit}

\begin{document}

\title{Derivation of the Born rule based on the minimal set of assumptions}

\author{A.~V.~Nenashev}
\email {nenashev@isp.nsc.ru}
\affiliation{Rzhanov Institute of Semiconductor Physics, 630090 Novosibirsk, Russia}
\affiliation{Novosibirsk State University, 630090 Novosibirsk, Russia}

\date{\today}

\begin{abstract}
The Born rule for probabilities of measurement results is deduced from the set 
of five assumptions. The assumptions state that: (a) the state vector fully 
determines the probabilities of all measurement results; (b) between 
measurements, any quantum system is governed by 
that part of standard quantum mechanics, which does not refer to measurements; 
(c) probabilities of measurement results obey the rules 
of the classical theory of probability; (d) no information transfer is 
possible without interaction; (e) if two spin-1/2 particles are in the entangled 
state $\sqrt{1-\lambda}\,\ket{\uparrow\uparrow}+\sqrt{\lambda}\,\ket{\downarrow\downarrow}$, 
and one of spins is measured by the Stern--Gerlach apparatus, then the state of 
the other spin after this measurement will be either $\ket{\uparrow}$ or 
$\ket{\downarrow}$, corresponding to the measurement result. No one of these 
assumptions can be omitted. 
The method of the derivation is based on Zurek's idea of environment-assisted 
invariance [PRL {\bf 90}, 120404 (2003)], though taken in rather modified form. 
Entanglement plays a crucial role in our approach (like in Zurek's one), so 
the probabilities are first considered 
in the case when the measured quantum system is 
entangled with some environment. 
Probabilities in pure states appear then as a particular case. 
The method of the present paper can be applied to both ideal and non-ideal 
measuring devices, irrespective to their constructions and principles of 
operation. 
\end{abstract}

\pacs{03.65.Ta,03.65.Yz,03.67.–a}

\maketitle

\section{Introduction}

The Born rule, stating that a probability is proportional to the squared absolute 
value of a complex amplitude, seems to stay apart from other fundamental concepts 
of quantum mechanics, as an independent postulate. 
On the other hand, no alternatives to the Born rule were invented so far. This 
suggests that the Born rule may actually be deduced from the rest of 
the quantum mechanics' corpus. There are many attempts to derive the Born rule, 
none of them is generally accepted by now. Born himself simply postulated his 
law for probabilities~\cite{Born1926}. Gleason~\cite{Gleason1957} proved a theorem about measures in a Hilbert 
space, that can be considered as a justification of the Born rule. Everett, in his 
famous paper devoted to the ``relative-state'' (or ``many-worlds'') interpretation 
of quantum mechanics~\cite{Everett1957}, argued that frequences of events have to obey the Born rule 
in any typical world. More references and a critical review of these efforts 
can be found in Ref.~\onlinecite{Schlosshauer2005}.

Recently, Zurek~\cite{Zurek2003} suggested a phenomenon of ``envariance'' 
(i.e.~environment-assisted invariance) as a tool to derive the Born rule. 
The key idea of envariance 
is that if two parts of a quantum system are entangled, then some unitary 
transformations acting on the \emph{first} subsystem can be completely undo by 
applying corresponding ``countertransformations'' to the \emph{second} subsystem. 
Then the state of the first subsystem is ``envariant'' under such transformations, 
which means that they do not change probabilities of any measurements' outcomes.
Here starts the way to the Born's rule. The following 
discussion~\cite{Barnum2003,Mohrhoff2004,Schlosshauer2005,Zurek2005,Herbut2006,Zurek2011,Herbut2012} 
had demonstrated 
a variety of opinions on how exactly can one arrive to the Born's rule from the envariance, 
and what additional assumptions one needs for that.

The aim of the present work is to make the arguments of Zurek more stronger. For this 
purpose, we introduce a minimal set of five assumptions (Assumptions 1--5 of 
Subsection~\ref{sub:st3-assump}) sufficient for the derivation of the Born rule. 
They are: three basic statements about quantum mechanics (Assumptions 1--3), 
the statement that there is no signalling without interaction (Assumption 4), 
and a proposition about the value of the state vector after the measurement (Assumption 5). 
No one of these assumptions can be omitted, as will be shown in Section~\ref{sec:di}. 

There are two features of the Zurek's derivation~\cite{Zurek2003,Zurek2005} and its versions 
by other authors~\cite{Barnum2003,Herbut2006,Herbut2012}, 
that can be considered as its restrictions. The first one is treating the probability 
$\mathcal{P}$ of the measurement result as a function of the state vector $\ket{\Psi}$ 
related to the system under measurement, and of the eigenvector $\ket{\varphi_n}$ 
related to the given ($n$-th) measurement result:
\begin{equation}
\mathcal{P} = \mathcal{P}\big( \ket{\Psi},\ket{\varphi_n} \big).
\end{equation}
There are at least two implicit assumptions in this: (a) that any measuring device is fully 
characterized (from the point of view of probabilities) by the set of eigenvectors of 
the corresponding observable; and (b) that if observables of two measuring devices have 
the same eigenvector $\ket{\varphi_n}$, then probabilities of obtanining the result 
corresponding to $\ket{\varphi_n}$ are the same for both devices. (These assumptions 
are also necessary for obtaining the Born rule from the Gleason's theorem.) 
These assumptions can be easily justified \emph{after} the Born rule has already been established, 
but it is not clear how to justify them without using the Born rule. 
Herbut~\cite{Herbut2012} avoids this problem by considering an ideal measurement of a 
special type. However our goal is to develop an approach suitable for \emph{any} 
measuring device. For this reason, we adhere to a more ``low-level'' view 
in the present paper---namely, we consider the probability $\mathcal{P}$ as a function of the 
state vector $\ket{\Psi}$, the measuring apparatus $\mathcal{A}$, and the number $n$ of the 
measurement result, without any references to observables as operators:
\begin{equation}
\mathcal{P} = \mathcal{P}_n\big( \ket{\Psi},\mathcal{A} \big).
\end{equation}
One more advantage of such ``low-level'' point of view is the possibility of considering 
a broader class of measuring devices: not only ``ideal'' devices described by operators, 
but also ones desribed by POVMs (positive operator-valued measures)~\cite{Peres1995}. 

The second problem is that Zurek's analysis is restricted by only those states in which 
the measured system is entangled with some other quantum system (an ``environment''), 
and the Schmidt decomposition of the state vector contains only eigenvectors of the measured 
observable. Pure states, as well as entangled states of general type, stay beyond the 
consideration. More complete analysis was performed by Herbut~\cite{Herbut2006}, 
but he was forced to make an additional assumption (see the ``sixth stipulation'' in 
Ref.~\onlinecite{Herbut2006}). We show that it is possible 
to proceed without such special assumption. 

We also got rid of the assumption of continuity of the probability, as a function of the 
state vector. The job of such assumption is made by Lemma~1 in the present paper.

Our approach is quite different in its form from Zurek's one, 
but the crucial role of entanglement in derivation of the Born rule is conserved. 
Namely, the most part of our paper is devoted to entangled states; pure states appear only as a 
particular case.

For the sake of simplicity, we concentrate on measurements of the spin 
degree of freedom of a spin-1/2 particle. This is not a restriction of our method---we 
show in Subsection~\ref{sub:pr4-coord} how to generalize the obtained results to 
other systems. 

The paper is organized as follows. Section~\ref{sec:st} contains a preliminary information:
statement of the problem, description of the Stern--Gerlach apparatus that serves as a 
``reference'' measuring device (Subsection~\ref{sub:st1-stern}), the language of quantum 
circuit diagrams (Subsection~\ref{sub:st2-notations}), and the list of assumptions 
(Subsection~\ref{sub:st3-assump}). 
Section~\ref{sec:pr} contains the derivation of the Born rule: a proof of linearity of 
functional dependence of the probability on the spin polarization vector
(Subsection~\ref{sub:pr2-polariz}---this is the main part of the paper), a proof of the Born rule 
for measuring the spin projection (Subsection~\ref{sub:pr3-spin}), and for measuring the 
particle coordinate (Subsection~\ref{sub:pr4-coord}). In Section~\ref{sec:di}, the question 
of necessity of each assumption is examined. Closing remarks are gathered in 
Section~\ref{sec:co}.

\section{Statement of the problem}
\label{sec:st}

Let us consider probabilities $\mathcal{P}_\uparrow(\mathcal{S},\mathcal{A})$ 
and $\mathcal{P}_\downarrow(\mathcal{S},\mathcal{A})$ of the outcomes 
$+1/2$ and $-1/2$ when the vertical projection of the spin is 
measured by means of some apparatus $\mathcal{A}$ for some spin-1/2 particle 
in a quantum state $\mathcal{S}$. There are two questions: 

{\bf Question 1.} Do these 
probabilites really depend on the apparatus $\mathcal{A}$, on its principle 
of operation, construction, etc.?

{\bf Question 2.} What is the mathematical expression for the functions 
$\mathcal{P}_\uparrow(\mathcal{S},\mathcal{A})$, 
$\mathcal{P}_\downarrow(\mathcal{S},\mathcal{A})$ with a fixed $\mathcal{A}$?

As we have no \emph{a priori} answer to the first question, we will proceed 
with the following strategy. First, we select some ``reference'' measurement 
device $\mathcal{A}_0$ for which it is possible to postulate an additional 
property (see Assumption 5 below), due to simplicity of the device. 
We choose the ``Stern--Gerlach apparatus'' described in Subsection~\ref{sub:st1-stern} as a 
``reference'' device. Then, in Subsection~\ref{sub:pr2-polariz}, 
we will consider thought experiments with 
a system of several entangled spins, when one spin is 
measured by the ``reference'' apparatus $\mathcal{A}_0$ and another spin is measured 
by some arbitrary apparatus $\mathcal{A}$. 
Analysis of these experiments gives the opportunity to use the knowledge about the apparatus 
$\mathcal{A}_0$ for making conclusions about the apparatus $\mathcal{A}$.
On this way we will manage to answer Questions~1 and 2.

In this Section, we describe the ``Stern--Gerlach apparatus'' for measuring the vertical 
component of the spin (Subsection~\ref{sub:st1-stern}), introduce notations for probabilities of 
different results of experiments (Subsection~\ref{sub:st2-notations}), 
make nesessary Assumptions and consider 
some consequences of them (Subsection~\ref{sub:st3-assump}). 
The detailed discussion of the set of 
Assumptions will be presented later, in Section~\ref{sec:di}.

\subsection{Stern--Gerlach apparatus}
\label{sub:st1-stern}

\begin{figure}
\includegraphics[width=\linewidth]{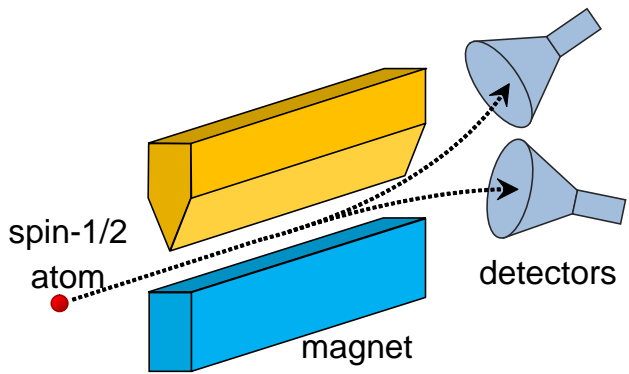}
\caption{Stern--Gerlach apparatus.}
\label{fig:stern}
\end{figure}

When a beam of spin-1/2 atoms (initially unpolarized) 
passes through the area of inhomogeneous magnetic field, it splits 
into two beams, each consists of spin-polarized 
atoms. This is the famous Stern--Gerlach experiment. 
One can use this effect also to measure the spin of a single atom, 
just by throwing it into the field  
and catching it at two possible exit trajectories 
(see Fig.~\ref{fig:stern}). The gradient of the 
magnetic field is chosen here to be parrallel to the vertical axis, therefore 
the apparatus is able to distinguish spin-up $\ket{\uparrow}$ 
and spin-down $\ket{\downarrow}$ states. We place two ideal detectors of atoms 
in the output of the apparatus, one of them catches everything that goes on 
the upper trajectory, another catches 
everything on the lower trajectory. When a detector catches an atom, 
it immediately reports this event. 
Later, this construction will be called the ``Stern--Gerlach apparatus''.

After passing the area of unhomogeneous magnetic field, the spin-up and 
spin-down parts of the atom's wavefunction are fully separated in space. 
This property of the Stern--Gerlach apparatus makes it
a proper candidate to the role of the ``reference'' measuring device. 
One can rely on this separation for justifying Assumption~5.

An obvious consequence of the spatial separation mentioned above is that, 
if the atom's spin is up, then 
no part of its wavefunction reaches the lower detector (assuming that the gradient 
of the $z$-component of the field is directed downwards). In other words, 
if the atom is in the pure spin-up state $\ket{\uparrow}$, it will  
for certain be caught by the upper detector. 
Analogously, the atom in the pure 
spin-down state $\ket{\downarrow}$ will  
be caught by the lower detector with the probability 1. 
Thus, we will say that the measurement outcome is ``spin up'' if the atom is 
caught by the upper detector; otherwise (if the atom 
is caught by the lower detector) the outcome is ``spin down''.

\subsection{Notations for probabilities}
\label{sub:st2-notations}

We will make extensive use of thought experimens consisting in unitary transformations and measurements on 
systems of several spins prepared initially in some states. The natural way for describing such 
experiments is the language of quantum circuit diagrams. This is a simple example of such diagram:
\[
  \ket{\uparrow} \; \Qcircuit @C=.5em @R=.8em { & \gate{U} & \measureD{\phantom{\uparrow}} } \; ,
\]
which means that the spin was prepared in the state $\ket{\uparrow}$, then the unitary operator $U$ was 
applied to the spin, and finally the measurement was performed under this spin. We reserve the sign 
$\Qcircuit @C=.5em { & \measureD{\phantom{\uparrow}} }$ for the measurement by the Stern--Gerlach apparatus; 
other measurements will be denoted as $\Qcircuit @C=.5em { & \measure{\phantom{xxx}} } \, $.

To denote \emph{probabilities} of measurement outcomes we will put the diagrams into square brackets, 
and indicate the particular outcome inside the measurement sign. For example, the following record:
\[
  \left[
  \ket{\uparrow} \; \Qcircuit @C=.5em @R=.8em { & \gate{U} & \measureD{\uparrow} } 
  \right]
\]
means the probability of the ``spin-up'' outcome when the spin, after preparing in the state $\ket{\uparrow}$ 
and applying the operator $U$, was measured by the Stern--Gerlach apparatus. The probability of the 
``spin-down'' outcome in the same experiment is denoted as
\[
  \left[
  \ket{\uparrow} \; \Qcircuit @C=.5em @R=.8em { & \gate{U} & \measureD{\downarrow} } 
  \right]
  .
\]
An analogous notation holds for the case of more than one spin. For example, this record:
\[
  \left[
  \ket{\rule[-2mm]{0mm}{5mm} \Psi} \begin{array}{c}\Qcircuit @C=.5em @R=.2em { 
  & \measureD{\uparrow} \\
  & \qw                 & \measureD{\downarrow} 
  }\end{array} 
  \right]
\]
denotes the probability that, for the system of two spins prepared in the state $\ket{\Psi}$, 
measurement of the first spin gives the result ``spin up'' \emph{and} successive measurement of the 
second spin gives the result ``spin down''.

\subsection{Assumptions}
\label{sub:st3-assump}

We start from three very basic assumptions.

\begin{itemize}
\item {\bf Assumption 1.} A state vector of a quantum system provides 
the full information about probabilities of all outcomes of measurements on this system 
and on its subsystems.
\item {\bf Assumption 2.} Between measurements, a quantum system obeys the rules of the 
non-measurement part of the quantum mechanics.
\item {\bf Assumption 3.} Probabilities of measurement results obey the rules of usual 
classical theory of probability.
\end{itemize}

Some consequences of these assumptions can be expressed in the language of quantum 
circuit diagrams introduced in the previous Subsection. First, the very statement that these 
notations for probabilities have any sense is a consequence of Assumption~1. 
Then, this Assumption guarantees that the probability has not been changed when an additional system 
is added into consideration, as in the following example:
\begin{equation}\label{a1-1}
  \left[
  \ket{\psi} \; \Qcircuit @C=.5em @R=.8em { & \measureD{\uparrow} } 
  \right]
  =
  \left[
  \quad\;\; 
  \begin{array}{c}\Qcircuit @C=.5em @R=.8em { 
  \lstick{\ket{\varphi}}
  & \qw                 & \qw \\ 
  \lstick{\ket{\psi}}
  & \measureD{\uparrow} 
  } \end{array}
  \right]
  ,
\end{equation}
where $\ket{\psi}$ and $\ket{\varphi}$ are arbitrary state vectors.

The second Assumption tells, in particular, that any quantum gates are described by unitary 
(norm-conserving) operators; that independent action of different quantum gates on different 
subsystems is expressed by the tensor product of corresponding operators.

The third Assumption allows for doing some arithmetics with probabilities. First, the ``spin-up'' 
and ``spin down'' outcomes form an exhaustive and mutually exclusive set of events, so the sum of 
their probabilities is unity:
\begin{equation}\label{a3-1}
  \left[
  \ket{\psi} \; \Qcircuit @C=.5em @R=.8em { & \measureD{\uparrow} } 
  \right]
  +
  \left[
  \ket{\psi} \; \Qcircuit @C=.5em @R=.8em { & \measureD{\downarrow} } 
  \right]
  =1.
\end{equation}

Let us consider a system of two spins prepared in some state $\ket{\Psi}$.
If the state of the first spin, after the measurement the second spin with the result $a$, 
is known to be $\ket{\varphi_a}$, then
\begin{equation}\label{a3-2}
  \left[
  \ket{\rule[-2mm]{0mm}{5mm} \Psi} \begin{array}{c}\Qcircuit @C=.5em @R=.2em { 
  & \qw                 & \measure{\rule{1.5mm}{0mm}b\rule{1.5mm}{0mm}} \\
  & \measureD{\vphantom\uparrow a} \\
  }\end{array} 
  \right]
  =
  \left[
  \ket{\rule[-2mm]{0mm}{5mm} \Psi} \begin{array}{c}\Qcircuit @C=.5em @R=.9em { 
  & \qw                 & \qw \\
  & \measureD{\vphantom\uparrow a} \\
  }\rule[-6.5mm]{0mm}{10mm} \end{array}  
  \right]
  \cdot
  \left[
  \quad\;\;\;\; \raisebox{11pt}{\Qcircuit @C=.5em @R=1.0em { 
  \lstick{\ket{\varphi_a}\!}
  & \measure{\rule{1.5mm}{0mm}b\rule{1.5mm}{0mm}} \\
  &  \\
  }} 
  \,\right]
  .
\end{equation}
Here the left hand side is the probability that the measurement of the 2nd spin 
gives the result $a$ 
and the subsequent measurement of the 1st spin gives the result $b$.
The first factor of the right hand side is the probability of the 
result $a$ when only the 2nd spin is measured;
the second factor plays the role of the \emph{conditional} probability of the 
result $b$ of the 1st spin's measurement 
\emph{provided that} the measurement of the 2nd spin gave the result $a$.
Therefore Eq.~(\ref{a3-2}) is simply an expression of the multiplicational rule 
for classical probabilities of successive events. 
(In this example, as well as in some examples below, the first spin is measured by 
some arbitrary device, and the second spin is measured by the Stern--Gerlach device.)

Another property of classical probability is causality: a probability of some event cannot 
depend on what happens \emph{after} the event, for example:
\begin{equation}\label{a3-3}
  \left[
  \ket{\rule[-2mm]{0mm}{5mm} \Psi} \begin{array}{c}\Qcircuit @C=.5em @R=.8em { 
  & \measure{\rule{1.5mm}{0mm}a\rule{1.5mm}{0mm}\rule[-0.5mm]{0mm}{2.5mm}} \\
  & \qw                 & \qw 
  }\rule[-6.5mm]{0mm}{10mm} \end{array}  
  \right]
  =
  \left[
  \ket{\rule[-2mm]{0mm}{5mm} \Psi} \begin{array}{c}\Qcircuit @C=.5em @R=.2em { 
  & \measure{\rule{1.5mm}{0mm}a\rule{1.5mm}{0mm}\rule[-0.5mm]{0mm}{2.5mm}} \\
  & \qw                 & \gate{U} & \qw
  }\end{array} 
  \right]
  ,
\end{equation}
where $U$ is any operation. 

Does Eq.~(\ref{a3-3}) remain valid if the operation $U$ is performed 
\emph{before} the measurement? Common sense says yes, because there is no interaction 
between the two spins, so any manipulations with the second spin cannot affect the probabilities 
of events that happens with the first spin. Let us formulate this statement in a more general form:

\begin{itemize}
\item {\bf Assumption 4.} No information transfer is possible between non-interacting systems.
\end{itemize}

If some manipulations with one system could cause change of probabilities associated with another 
system, then one could use them for signalling without interaction. So, ``no information transfer'' 
implies also that any action on one system cannot change probabilities of other system's measurement 
results. For example,
\begin{equation}\label{a4-1}
  \left[
  \ket{\rule[-2mm]{0mm}{5mm} \Psi} \begin{array}{c}\Qcircuit @C=.5em @R=.8em { 
  & \measure{\rule{1.5mm}{0mm}a\rule{1.5mm}{0mm}\rule[-0.5mm]{0mm}{2.5mm}} \\
  & \qw                 & \qw 
  }\rule[-6.5mm]{0mm}{10mm} \end{array}  
  \right]
  =
  \left[
  \ket{\rule[-2mm]{0mm}{5mm} \Psi} \begin{array}{c}\Qcircuit @C=.5em @R=.2em { 
  & \qw      & \measure{\rule{1.5mm}{0mm}a\rule{1.5mm}{0mm}\rule[-0.5mm]{0mm}{2.5mm}} \\
  & \gate{U} & \qw                 & \qw
  }\end{array} 
  \right]
  .
\end{equation}
Also the \emph{measurement} of one system does not change probabilities for another system, 
if the measurement result is unknown:
\begin{equation}\label{a4-2}
  \left[
  \ket{\rule[-2mm]{0mm}{5mm} \Psi} \begin{array}{c}\Qcircuit @C=.5em @R=.8em { 
  & \measure{\rule{1.5mm}{0mm}a\rule{1.5mm}{0mm}\rule[-0.5mm]{0mm}{2.5mm}} \\
  & \qw                 & \qw 
  }\rule[-6.5mm]{0mm}{10mm} \end{array}  
  \right]
  =
  \left[
  \ket{\rule[-2mm]{0mm}{5mm} \Psi} \begin{array}{c}\Qcircuit @C=.5em @R=.2em { 
  & \qw                           & \measure{\rule{1.5mm}{0mm}a\rule{1.5mm}{0mm}\rule[-0.5mm]{0mm}{2.5mm}} \\
  & \measureD{\phantom{\uparrow}} 
  }\end{array} 
  \right]
  .
\end{equation}
According to the additive rule of the probability theory, the diagram in the right hand side can be 
expanded as
\begin{multline}\label{a4-3}
   \left[
  \ket{\rule[-2mm]{0mm}{5mm} \Psi} \begin{array}{c}\Qcircuit @C=.5em @R=.2em { 
  & \qw                           & \measure{\rule{1.5mm}{0mm}a\rule{1.5mm}{0mm}\rule[-0.5mm]{0mm}{2.5mm}} \\
  & \measureD{\phantom{\uparrow}} 
  }\end{array} 
  \right]
  =
  \\
  = 
  \left[
  \ket{\rule[-2mm]{0mm}{5mm} \Psi} \begin{array}{c}\Qcircuit @C=.5em @R=.2em { 
  & \qw                 & \measure{\rule{1.5mm}{0mm}a\rule{1.5mm}{0mm}\rule[-0.5mm]{0mm}{2.5mm}} \\
  & \measureD{\uparrow} 
  }\end{array} 
  \right]
  +
  \left[
  \ket{\rule[-2mm]{0mm}{5mm} \Psi} \begin{array}{c}\Qcircuit @C=.5em @R=.2em { 
  & \qw                   & \measure{\rule{1.5mm}{0mm}a\rule{1.5mm}{0mm}\rule[-0.5mm]{0mm}{2.5mm}} \\
  & \measureD{\downarrow} 
  }\end{array} 
  \right]
  .
\end{multline}

Now let us postulate one property of the Stern--Gerlach apparatus. 
When the atom is eaten by a detector of the Stern--Gerlach apparatus, its state apparently 
cannot be further traced. 
But, if a system of \emph{several} spins is under consideration, 
it is important to know the state of other spins after measurement of one spin.
Let us consider the simplest case of two spins in the state 
\begin{equation}\label{a5-0}
  \ket{S_\lambda} = \sqrt{1-\lambda}\,\ket{\uparrow\uparrow}+\sqrt{\lambda}\,\ket{\downarrow\downarrow},
\end{equation}
where $\lambda\in[0,1]$.
It is natural to suppose that, if one spin is found to be up, the other will be also up; 
otherwise (one spin down), the other spin will be down. So we accept the following statement.

\begin{itemize}
\item {\bf Assumption 5.} If a system of two spin-1/2 particles
was in the state 
$\sqrt{1-\lambda}\,\ket{\uparrow\uparrow}+\sqrt{\lambda}\,\ket{\downarrow\downarrow}$, 
where $0\leq\lambda\leq1$, and one of the spins is measured by the Stern--Gerlach apparatus,
then the state of the other spin after the measurement will be $\ket{\uparrow}$ in the case of 
the ``spin-up'' measurement result, and $\ket{\downarrow}$ in the case of the ``spin-down'' result.
\end{itemize}

Though Assumption~5 looks like collapse, it also makes sense in no-collapse interpretations 
of quantum mechanics. In the Everettian interpretation~\cite{Everett1957}, states $\ket{\uparrow}$ 
and $\ket{\downarrow}$ of the remaining spin after the measurement have the meaning of states 
\emph{relative to} the measurement results $\uparrow$ and $\downarrow$.

Using Eq.~(\ref{a3-2}), one can express Assumption~5 in the language of quantum circuit diagrams:
\begin{gather}
\label{a5-1}
  \left[
  \ket{\rule[-2mm]{0mm}{5mm} S_\lambda} \begin{array}{c}\Qcircuit @C=.5em @R=.2em { 
  & \qw                 & \measure{\rule{1.5mm}{0mm}a\rule{1.5mm}{0mm}\rule[-0.5mm]{0mm}{2.5mm}} \\
  & \measureD{\uparrow} \\
  }\end{array} 
  \right]
  =
  \left[
  \ket{\rule[-2mm]{0mm}{5mm} S_\lambda} \begin{array}{c}\Qcircuit @C=.5em @R=.9em { 
  & \qw                 & \qw \\
  & \measureD{\uparrow} \\
  }\rule[-6.5mm]{0mm}{10mm} \end{array}  
  \right]
  \cdot
  \left[
  \quad\;\; \raisebox{10pt}{\Qcircuit @C=.5em @R=1.0em { 
  \lstick{\ket{\uparrow}\!}
  & \measure{\rule{1.5mm}{0mm}a\rule{1.5mm}{0mm}\rule[-0.5mm]{0mm}{2.5mm}} \\
  &  \\
  }} 
  \,\right]
  ,
\\
\label{a5-2}
  \left[
  \ket{\rule[-2mm]{0mm}{5mm} S_\lambda} \begin{array}{c}\Qcircuit @C=.5em @R=.2em { 
  & \qw                 & \measure{\rule{1.5mm}{0mm}a\rule{1.5mm}{0mm}\rule[-0.5mm]{0mm}{2.5mm}} \\
  & \measureD{\downarrow} \\
  }\end{array} 
  \right]
  =
  \left[
  \ket{\rule[-2mm]{0mm}{5mm} S_\lambda} \begin{array}{c}\Qcircuit @C=.5em @R=.9em { 
  & \qw                 & \qw \\
  & \measureD{\downarrow} \\
  }\rule[-6.5mm]{0mm}{10mm} \end{array}  
  \right]
  \cdot
  \left[
  \quad\;\; \raisebox{10pt}{\Qcircuit @C=.5em @R=1.0em { 
  \lstick{\ket{\downarrow}\!}
  & \measure{\rule{1.5mm}{0mm}a\rule{1.5mm}{0mm}\rule[-0.5mm]{0mm}{2.5mm}} \\
  &  \\
  }} 
  \,\right]
  ,
\end{gather}
where 
$\Qcircuit @C=.5em @R=1.0em { & \measure{\rule{1.5mm}{0mm}a\rule{1.5mm}{0mm}\rule[-0.5mm]{0mm}{2.5mm}} }$ 
can be any measurement, or even a combination of unitary transformations and measurements. 
Then, substituting Eqs.~(\ref{a4-3}),(\ref{a5-1}),(\ref{a5-2}) into (\ref{a4-2}), one can get 
the following relation:
\begin{multline}\label{a5-3}
  \left[
  \ket{\rule[-2mm]{0mm}{5mm} S_\lambda} \begin{array}{c}\Qcircuit @C=.5em @R=.8em { 
  & \measure{\rule{1.5mm}{0mm}a\rule{1.5mm}{0mm}\rule[-0.5mm]{0mm}{2.5mm}} \\
  & \qw                 & \qw 
  }\rule[-6.5mm]{0mm}{10mm} \end{array}  
  \right]
  =
  \left[
  \ket{\rule[-2mm]{0mm}{5mm} S_\lambda} \begin{array}{c}\Qcircuit @C=.5em @R=.9em { 
  & \qw                 & \qw \\
  & \measureD{\uparrow} \\
  }\rule[-6.5mm]{0mm}{10mm} \end{array}  
  \right]
  \cdot
  \left[
  \quad\;\; \raisebox{10pt}{\Qcircuit @C=.5em @R=1.0em { 
  \lstick{\ket{\uparrow}\!}
  & \measure{\rule{1.5mm}{0mm}a\rule{1.5mm}{0mm}\rule[-0.5mm]{0mm}{2.5mm}} \\
  &  \\
  }} 
  \,\right]
  \\
  +
  \left[
  \ket{\rule[-2mm]{0mm}{5mm} S_\lambda} \begin{array}{c}\Qcircuit @C=.5em @R=.9em { 
  & \qw                 & \qw \\
  & \measureD{\downarrow} \\
  }\rule[-6.5mm]{0mm}{10mm} \end{array}  
  \right]
  \cdot
  \left[
  \quad\;\; \raisebox{10pt}{\Qcircuit @C=.5em @R=1.0em { 
  \lstick{\ket{\downarrow}\!}
  & \measure{\rule{1.5mm}{0mm}a\rule{1.5mm}{0mm}\rule[-0.5mm]{0mm}{2.5mm}} \\
  &  \\
  }} 
  \,\right]
  .
\end{multline}
This equation will be used in Subsection~\ref{sub:pr2-polariz}.

The set of five Assumptions listed here provides the minimal background for 
derivation of the Born rule.

\section{Proof of the Born rule}
\label{sec:pr}

\subsection{Probability and spin polarizaton}
\label{sub:pr2-polariz}

\newcommand{\cgate}[1]{*+<.6em>{#1} \POS ="i","i"+UR;"i"+UL **\dir{-};"i"+DL **\dir{-};"i"+DR **\dir{-};"i"+UR **\dir{-},"i" \cw}

We will consider an arbitrary measuring device, treating it as a black box which accepts 
a spin-1/2 particle on the input, and gives some classical information on the output. 
It is supposed that the measuring device interacts only with the spin degrees of freedom 
of the particle under measurement. For simplicity, we consider only one bit of information 
on the output---namely, we will say that the device either ``clicks'' of ``not clicks'' 
when a particle comes into it. This simplification does not lead to any loss of generality when 
we are interested in probabilities of measurement outcomes. Indeed, one can select some  
particular outcome and consider the appearance of this outcome as ``clicking''.

Let $\mathcal{P}_{\text{click}}\bigl(\ket{\mathbf{\Psi}}\bigr)$ 
be the probability of ``clicking'' the detector when it interacts with the spin:
\begin{equation}
  \mathcal{P}_{\text{click}}\bigl(\ket{\mathbf{\Psi}}\bigr) 
  = 
  \left[
  \ket{\rule[-2mm]{0mm}{5mm} \mathbf{\Psi}} \begin{array}{c}\Qcircuit @C=.5em @R=.8em { 
  & \measure{\text{click}} \\
  & \cw                    & \cw 
  }\rule[-6.5mm]{0mm}{10mm} \end{array}  
  \right] 
  ,
\end{equation}
where $\ket{\mathbf{\Psi}}$ is some state vector of the composite system ``spin + environment''.
An environment is any quantum system separated from the measured particle, or even the whole 
Universe except the particle and the measuring device.
The single wire $(\Qcircuit @C=1em {&\qw})$ in this diagram refers to the spin, 
and the double wire $(\Qcircuit @C=1em {&\cw})$ refers to the environment.

Any state vector of this composite system can be represented in the form of Schmidt decomposition:
\begin{equation}
  \ket{\mathbf{\Psi}} = c_1\;\ket{a_1}\ket{\mathbf{b}_1} + c_2\;\ket{a_2}\ket{\mathbf{b}_2} ,
\end{equation}
where $c_1$ and $c_2$ are non-negative real numbers; $\ket{a_1}$ and $\ket{a_2}$ are two mutually 
orthogonal unit vectors of the spin; $\ket{\mathbf{b}_1}$ and $\ket{\mathbf{b}_2}$ are two mutually 
orthogonal unit vectors of the environment:
\begin{equation}
  c_1\geq0, \;\; c_2\geq0, \quad 
  \langle a_k|a_l \rangle = \langle \mathbf{b}_k|\mathbf{b}_l \rangle = \delta_{kl}.
\end{equation}

Let two vectors $\ket{\mathbf{\Psi}'}$ and $\ket{\mathbf{\Psi}''}$ have Schmidt decompositions 
with the same $c_1$, $c_2$, $\ket{a_1}$ and $\ket{a_2}$: 
\begin{align}
\label{pr2-psip}
  \ket{\mathbf{\Psi}'}  &= c_1\;\ket{a_1}\ket{\mathbf{b}'_1}  + c_2\;\ket{a_2}\ket{\mathbf{b}'_2} , \\
\label{pr2-psipp}
  \ket{\mathbf{\Psi}''} &= c_1\;\ket{a_1}\ket{\mathbf{b}''_1} + c_2\;\ket{a_2}\ket{\mathbf{b}''_2} .
\end{align}
Then, there exists a unitary transformation $\mathbf{U}$ acting on the environment degrees of freedom, 
which maps the set $\{\ket{\mathbf{b}'_1},\ket{\mathbf{b}'_2}\}$ of unit vectors into 
the set $\{\ket{\mathbf{b}''_1},\ket{\mathbf{b}''_2}\}$:
\begin{equation}
\label{pr2-u}
  \mathbf{U}\,\ket{\mathbf{b}'_1} = \ket{\mathbf{b}''_1}, \quad
  \mathbf{U}\,\ket{\mathbf{b}'_2} = \ket{\mathbf{b}''_2}.
\end{equation}
It follows from Eqs.~(\ref{pr2-psip})--(\ref{pr2-u}) that
\begin{equation}
\label{pr2-u-property}
  \ket{\rule[-2mm]{0mm}{5mm} \mathbf{\Psi}'} 
  \begin{array}{c}\Qcircuit @C=.5em @R=.8em { 
  & \qw                & \qw \\
  & \cgate{\mathbf{U}} & \cw 
  } \end{array}  
  \quad = \quad
  \ket{\rule[-2mm]{0mm}{5mm} \mathbf{\Psi}''} .
\end{equation}
Using this transformation, it is easy to prove that
\begin{equation}
\label{pr2-same-pclick}
  \mathcal{P}_{\text{click}}\bigl(\ket{\mathbf{\Psi}'}\bigr) = 
  \mathcal{P}_{\text{click}}\bigl(\ket{\mathbf{\Psi}''}\bigr) .
\end{equation}
Indeed, let us apply Eq.~(\ref{a4-1}), and then Eq.~(\ref{pr2-u-property}):
\begin{multline*}
  \mathcal{P}_{\text{click}}\bigl(\ket{\mathbf{\Psi}'}\bigr) 
  \equiv 
  \left[
  \ket{\rule[-2mm]{0mm}{5mm} \mathbf{\Psi}'} \begin{array}{c}\Qcircuit @C=.5em @R=.8em { 
  & \measure{\text{click}} \\
  & \cw                    & \cw 
  }\rule[-6.5mm]{0mm}{10mm} \end{array}  
  \right] 
  \\
  =
  \left[
  \ket{\rule[-2mm]{0mm}{5mm} \mathbf{\Psi}'} \begin{array}{c}\Qcircuit @C=.5em @R=.2em { 
  & \qw                & \measure{\text{click}} \\
  & \cgate{\mathbf{U}} & \cw                    & \cw
  }\rule[-6.5mm]{0mm}{10mm} \end{array}  
  \right] 
  \\
  = 
  \left[
  \ket{\rule[-2mm]{0mm}{5mm} \mathbf{\Psi}''} \begin{array}{c}\Qcircuit @C=.5em @R=.8em { 
  & \measure{\text{click}} \\
  & \cw                    & \cw 
  }\rule[-6.5mm]{0mm}{10mm} \end{array}  
  \right] 
  \equiv
  \mathcal{P}_{\text{click}}\bigl(\ket{\mathbf{\Psi}''}\bigr) 
  .
\end{multline*}

Now we introduce the spin polarization vector $\mathbf{p}=(p_x,p_y,p_z)$ 
defined (for the normalized state vector $\ket{\mathbf{\Psi}}$) as follows:
\begin{equation}
\label{pr2-p-def}
  \mathbf{p}\bigl(\ket{\mathbf{\Psi}}\bigr) = 
  \langle \mathbf{\Psi} | \hat{\boldsymbol{\sigma}} 
  \otimes \mathbb{I}_\mathcal{E} | \mathbf{\Psi} \rangle \,,
\end{equation}
where $\hat{\boldsymbol{\sigma}}=(\hat\sigma_x,\hat\sigma_y,\hat\sigma_z)$ 
is the vector of Pauli matrices, and $\mathbb{I}_\mathcal{E}$ 
is the identity matrix in the environment state space. 
The set of possible values of $\mathbf{p}$ is the ball $|\mathbf{p}|\leq1$ (the Bloch ball). 

One can easily see that states $\ket{\mathbf{\Psi}'}$ and $\ket{\mathbf{\Psi}''}$ 
defined by Eqs.~(\ref{pr2-psip}) and (\ref{pr2-psipp}) have the same spin polarization. 
The converse statement is aslo true: if two normalized state vectors $\ket{\mathbf{\Psi}_1}$ and 
$\ket{\mathbf{\Psi}_2}$ have the same spin polarization, then 
there exist Schmidt decompositions for these vectors with the same $c_1$, $c_2$, 
$\ket{a_1}$, and $\ket{a_2}$. Taking Eq.~(\ref{pr2-same-pclick}) into account, 
we arrive to the conclusion that,
if two state vectors have equal spin polarization, they have equal 
probability of ``clicking'' the detector:
\begin{equation}
\label{pr2-function0}
  \mathbf{p}\bigl(\ket{\mathbf{\Psi}_1}\bigr) =
  \mathbf{p}\bigl(\ket{\mathbf{\Psi}_2}\bigr) 
  \;\; \Longrightarrow \;\;
  \mathcal{P}_{\text{click}}\bigl(\ket{\mathbf{\Psi}_1}\bigr) = 
  \mathcal{P}_{\text{click}}\bigl(\ket{\mathbf{\Psi}_2}\bigr) .
\end{equation}
In other words, $\mathcal{P}_{\text{click}}\bigl(\ket{\mathbf{\Psi}}\bigr)$ 
is a function of $\mathbf{p}\bigl(\ket{\mathbf{\Psi}}\bigr)$. 
We denote this function as $F_{\text{click}}(\mathbf{p})$:
\begin{equation}
\label{pr2-function1}
  \mathcal{P}_{\text{click}}\bigl(\ket{\mathbf{\Psi}}\bigr) = 
  F_{\text{click}}\bigl(\mathbf{p}(\ket{\mathbf{\Psi}})\bigr) .
\end{equation}

Then we will formulate and prove three statements about the function $F_{\text{click}}(\mathbf{p})$.

\ 

{\bf Lemma 1.} For any two vectors $\mathbf{p}_0$ and $\mathbf{p}_1$ in the Bloch ball 
$(|\mathbf{p}_0|\leq1,|\mathbf{p}_1|\leq1)$ and any number $\lambda\in[0,1]$, 
the value $F_{\text{click}}\bigl((1-\lambda)\mathbf{p}_0+\lambda\mathbf{p}_1\bigr)$ 
lies between $F_{\text{click}}(\mathbf{p}_0)$ and $F_{\text{click}}(\mathbf{p}_1)$:
\begin{align*}
  F_{\text{click}}(\mathbf{p}_0) &\leq 
  F_{\text{click}}\bigl((1-\lambda)\mathbf{p}_0+\lambda\mathbf{p}_1\bigr) \leq 
  F_{\text{click}}(\mathbf{p}_1) \;\; \text{or} \\
  F_{\text{click}}(\mathbf{p}_1) &\leq 
  F_{\text{click}}\bigl((1-\lambda)\mathbf{p}_0+\lambda\mathbf{p}_1\bigr) \leq 
  F_{\text{click}}(\mathbf{p}_0) .
\end{align*}

{\bf Lemma 2.} For any two vectors $\mathbf{p}_0$ and $\mathbf{p}_1$ in the Bloch ball 
$(|\mathbf{p}_0|\leq1,|\mathbf{p}_1|\leq1)$ 
\begin{equation*}
  F_{\text{click}}\left( \frac{\mathbf{p}_0+\mathbf{p}_1}{2} \right) = 
  \frac{ F_{\text{click}}(\mathbf{p}_0) + F_{\text{click}}(\mathbf{p}_1) }{2} \, .
\end{equation*}

{\bf Lemma 3.} For any two vectors $\mathbf{p}_0$ and $\mathbf{p}_1$ in the Bloch ball 
$(|\mathbf{p}_0|\leq1,|\mathbf{p}_1|\leq1)$ and any number $\lambda\in[0,1]$,
\begin{equation*}
  F_{\text{click}}\bigl((1-\lambda)\mathbf{p}_0+\lambda\mathbf{p}_1\bigr) = 
  (1-\lambda)F_{\text{click}}(\mathbf{p}_0) + \lambda\, F_{\text{click}}(\mathbf{p}_1) .
\end{equation*}

{\bf Proof of Lemma~1.} It is always possible to find such unit vectors 
$\ket{\psi_0}$ and $\ket{\psi_1}$ in a state space of two spins that 
\begin{equation}
  \mathbf{p}\bigl(\ket{\psi_0}\bigr) = \mathbf{p}_0, \quad 
  \mathbf{p}\bigl(\ket{\psi_1}\bigr) = \mathbf{p}_1,
\end{equation}
where the second spin plays the role of the environment. 
According to Eq.~(\ref{pr2-function1}),
\begin{gather}
\label{pr2-l1-proof1}
  \mathcal{P}_{\text{click}}\bigl(\ket{\psi_0}\bigr) = 
  F_{\text{click}}\bigl(\mathbf{p}_0\bigr) , \\
\label{pr2-l1-proof2}
  \mathcal{P}_{\text{click}}\bigl(\ket{\psi_1}\bigr) = 
  F_{\text{click}}\bigl(\mathbf{p}_1\bigr) .
\end{gather}
Now, we add two more spins into the environment, and construct 
a four-spin state vector $\ket{\Psi_\lambda}$ as follows:
\begin{equation}
  \ket{\Psi_\lambda} = 
  \sqrt{1-\lambda}  \,\ket{\psi_0}\ket{\uparrow\uparrow} +
  \sqrt{\lambda}\,\ket{\psi_1}\ket{\downarrow\downarrow} .
\end{equation}
Let us calculate the spin polarization for the state $\ket{\Psi_\lambda}$: 
\begin{multline*}
  \mathbf{p}\bigl(\ket{\Psi_\lambda}\bigr) 
  = 
  \langle \Psi_\lambda | \hat{\boldsymbol{\sigma}} 
  \otimes \mathbb{I}_3 | \Psi_\lambda \rangle 
  \\
  =
  (1-\lambda)\bra{\psi_0} \hat{\boldsymbol{\sigma}} 
    \otimes \mathbb{I}_1 \ket{\psi_0} + 
  \lambda\bra{\psi_1} \hat{\boldsymbol{\sigma}} 
    \otimes \mathbb{I}_1 \ket{\psi_1}  
  \\
  = 
  (1-\lambda)\mathbf{p}_0+\lambda\mathbf{p}_1 
\end{multline*}
($\mathbb{I}_1$ and $\mathbb{I}_3$ being identity matrices for one spin and three spins).
Therefore, according to Eq.~(\ref{pr2-function1}),
\begin{equation}
\label{pr2-l1-proof3}
  \mathcal{P}_{\text{click}}\bigl(\ket{\Psi_\lambda}\bigr) = 
  F_{\text{click}}\bigl((1-\lambda)\mathbf{p}_0+\lambda\mathbf{p}_1\bigr) .
\end{equation}

On the other hand, it is possible to evaluate the probability 
$\mathcal{P}_{\text{click}}\bigl(\ket{\Psi_\lambda}\bigr)$
using the fact that (according to Assumption~4) applying the Stern--Gerlach measurenent 
$\Qcircuit @C=.5em { & \measureD{\phantom{\uparrow}} }$
to the 4th spin does not change this probability. 
To do so, we first note that three-spin state vectors 
$\ket{\psi_0}\ket{\uparrow}$ and $\ket{\psi_1}\ket{\downarrow}$ 
are mutually orthogonal unit vectors. Consequently it is possible to map them into 
the basis vectors $\ket{\uparrow\uparrow\uparrow}$ and $\ket{\uparrow\uparrow\downarrow}$ 
by some unitary transformation $V$:
\begin{equation} 
\label{pr2-l1-v}
  V\,\ket{\psi_0}\ket{\uparrow}   = \ket{\uparrow\uparrow\uparrow}, \quad 
  V\,\ket{\psi_1}\ket{\downarrow} = \ket{\uparrow\uparrow\downarrow} .
\end{equation}
The inverse transformation $V^{-1}$ maps vectors $\ket{\uparrow\uparrow\uparrow}$ and 
$\ket{\uparrow\uparrow\downarrow}$ into $\ket{\psi_0}\ket{\uparrow}$ and 
$\ket{\psi_1}\ket{\downarrow}$:
\begin{equation} 
\label{pr2-l1-v-inv}
  V^{-1}\,\ket{\uparrow\uparrow\uparrow}   = \ket{\psi_0}\ket{\uparrow}, \quad 
  V^{-1}\,\ket{\uparrow\uparrow\downarrow} = \ket{\psi_1}\ket{\downarrow} .
\end{equation}
Let us apply the operator $V$ to 1st, 2nd and 3rd spins of a system of four spins 
in the state $\ket{\Psi_\lambda}$:
\begin{multline}
\label{pr2-l1-v-property0}
  V \ket{\Psi_\lambda} 
  = 
  \sqrt{1-\lambda} \left( V \ket{\psi_0}\ket{\uparrow}   \vphantom{\sqrt{t}}\right) \ket{\uparrow} + 
  \sqrt{\lambda}   \left( V \ket{\psi_1}\ket{\downarrow} \vphantom{\sqrt{t}}\right) \ket{\downarrow} \\ 
  = 
  \sqrt{1-\lambda}\,\ket{\uparrow\uparrow\uparrow\uparrow} + 
  \sqrt{\lambda}  \,\ket{\uparrow\uparrow\downarrow\downarrow} \\
  = 
  \ket{\uparrow\uparrow} 
  \left( \sqrt{1-\lambda}\,\ket{\uparrow\uparrow} + 
  \sqrt{\lambda}\,\ket{\downarrow\downarrow} \right) 
  = 
  \ket{\uparrow\uparrow}\ket{S_\lambda} ,
\end{multline}
where the two-spin vector $\ket{S_\lambda}$ is defined by Eq.~(\ref{a5-0}). The 
transformation rule~(\ref{pr2-l1-v-property0}) can be expressed also in a symbolic form:
\begin{equation} 
\label{pr2-l1-v-property}
  \Biggl\vert\Psi_\lambda\!\Biggr\rangle \;\;
  \raisebox{16pt}{\Qcircuit @C=.5em @R=0em @!R {
  & \multigate{2}{V} & \qw \\
  & \ghost{V}        & \qw \\
  & \ghost{V}        & \qw \\
  & \qw                       & \qw 
  }}
  \quad \equiv \rule{12mm}{0mm}
  \raisebox{16pt}{\Qcircuit @C=1.0em @R=0.9em @!R {
  \lstick{\ket{\uparrow}}
  & \qw \\
  \lstick{\ket{\uparrow}\,}
  & \qw \\
  \lstick{ \rule{0mm}{7mm} \ket{S_\lambda} \! \left\{ \rule{0mm}{3.5mm} \right. \!\!\!}
  & \qw \\
  & \qw 
  }}
\end{equation}
Now everything is ready for evaluation the quantity 
$\mathcal{P}_{\text{click}}\bigl(\ket{\Psi_\lambda}\bigr)$ 
(numbers above equality signs denote which rule or equation is used at the given step):
\begin{multline}
\label{pr2-l1-proof4}
  \mathcal{P}_{\text{click}}\bigl(\ket{\Psi_\lambda}\bigr) 
  \equiv 
  \left[ 
  \Biggl\vert\Psi_\lambda\!\Biggr\rangle \; \rule{0mm}{8mm}
  \raisebox{16pt}{\Qcircuit @C=.5em @R=-0.4em @!R {
  & \measure{\text{click}} \\
  & \qw                    & \qw \\
  & \qw                    & \qw \\
  & \qw                    & \qw \\
  }} \right]
  =
  \\
  =
  \left[ 
  \Biggl\vert\Psi_\lambda\!\Biggr\rangle \; \rule{0mm}{8mm}
  \raisebox{16pt}{\Qcircuit @C=.5em @R=-0.4em @!R {
  & \multigate{2}{V} & \multigate{2}{V^{-1}} & \measure{\text{click}} \\
  & \ghost{V}        & \ghost{V^{-1}}        & \qw                    & \qw \\
  & \ghost{V}        & \ghost{V^{-1}}        & \qw                    & \qw \\
  & \qw              & \qw                   & \qw                    & \qw \\
  }} \right]
  \\
  \stackrel{(\ref{pr2-l1-v-property})}{=}
  \left[ 
  \rule{9mm}{0mm} \rule{0mm}{8mm}
  \raisebox{16pt}{\Qcircuit @C=.5em @R=-0.4em @!R {
  \lstick{\ket{\uparrow}}
  & \multigate{2}{V^{-1}} & \measure{\text{click}} \\
  \lstick{\ket{\uparrow}\,}
  & \ghost{V^{-1}}        & \qw                    & \qw \\
  \lstick{ \rule{0mm}{7mm} \ket{S_\lambda} \! \left\{ \rule{0mm}{3.5mm} \right. \!\!\!}
  & \ghost{V^{-1}}        & \qw                    & \qw \\
  & \qw                   & \qw                    & \qw \\
  }} \right]
  \\
  \stackrel{(\ref{a5-3})}{=}
  \left[ 
  \rule{9mm}{0mm} \rule{0mm}{8mm}
  \raisebox{16pt}{\Qcircuit @C=.5em @R=-0.4em @!R {
  \lstick{\ket{\uparrow}}
  & \qw                   & \qw \\
  \lstick{\ket{\uparrow}\,}
  & \qw                   & \qw \\
  \lstick{ \rule{0mm}{7mm} \ket{S_\lambda} \! \left\{ \rule{0mm}{3.5mm} \right. \!\!\!}
  & \qw                   & \qw \\
  & \measureD{\uparrow} \\
  }} \right]
  \cdot
  \left[ 
  \rule{5mm}{0mm} \rule{0mm}{8mm}
  \raisebox{16pt}{\Qcircuit @C=.5em @R=-0.3em @!R {
  \lstick{\ket{\uparrow}\!\!}
  & \multigate{2}{V^{-1}} & \measure{\text{click}} \\
  \lstick{\ket{\uparrow}\!}
  & \ghost{V^{-1}}        & \qw                    & \qw \\
  \lstick{\ket{\uparrow}\!\!}
  & \ghost{V^{-1}}        & \qw                    & \qw \\
  & \\
  }} \right]
  \\
  \;\; + \,
  \left[ 
  \rule{9mm}{0mm} \rule{0mm}{8mm}
  \raisebox{16pt}{\Qcircuit @C=.5em @R=-0.4em @!R {
  \lstick{\ket{\uparrow}}
  & \qw                   & \qw \\
  \lstick{\ket{\uparrow}\,}
  & \qw                   & \qw \\
  \lstick{ \rule{0mm}{7mm} \ket{S_\lambda} \! \left\{ \rule{0mm}{3.5mm} \right. \!\!\!}
  & \qw                   & \qw \\
  & \measureD{\downarrow} \\
  }} \right]
  \cdot
  \left[ 
  \rule{5mm}{0mm} \rule{0mm}{8mm}
  \raisebox{16pt}{\Qcircuit @C=.5em @R=-0.3em @!R {
  \lstick{\ket{\uparrow}\!\!}
  & \multigate{2}{V^{-1}} & \measure{\text{click}} \\
  \lstick{\ket{\uparrow}\!}
  & \ghost{V^{-1}}        & \qw                    & \qw \\
  \lstick{\ket{\downarrow}\!\!}
  & \ghost{V^{-1}}        & \qw                    & \qw \\
  & \\
  }} \right]
  \\
  \stackrel{(\ref{pr2-l1-alambda},\ref{a3-1})}{=}
  a_\lambda
  \left[ 
  \rule{5mm}{0mm} 
  \raisebox{11pt}{\Qcircuit @C=.5em @R=-0.3em @!R {
  \lstick{\ket{\uparrow}\!\!}
  & \multigate{2}{V^{-1}} & \measure{\text{click}} \\
  \lstick{\ket{\uparrow}\!}
  & \ghost{V^{-1}}        & \qw                    & \qw \\
  \lstick{\ket{\uparrow}\!\!}
  & \ghost{V^{-1}}        & \qw                    & \qw \\
  }} \right]
  \\
  + (1-a_\lambda)
  \left[ 
  \rule{5mm}{0mm} 
  \raisebox{11pt}{\Qcircuit @C=.5em @R=-0.3em @!R {
  \lstick{\ket{\uparrow}\!\!}
  & \multigate{2}{V^{-1}} & \measure{\text{click}} \\
  \lstick{\ket{\uparrow}\!}
  & \ghost{V^{-1}}        & \qw                    & \qw \\
  \lstick{\ket{\downarrow}\!\!}
  & \ghost{V^{-1}}        & \qw                    & \qw \\
  }} \right]
  \\
  \stackrel{(\ref{pr2-l1-v-inv})}{=}
  a_\lambda
  \left[ 
  \rule{9mm}{0mm} 
  \raisebox{11pt}{\Qcircuit @C=.5em @R=-0.3em @!R {
  & \measure{\text{click}} \\
  \lstick{ \rule[-6mm]{0mm}{7mm} \ket{\psi_0} \! \left\{ \rule{0mm}{3.5mm} \right. \!\!\!}
  & \qw                    & \qw \\
  \lstick{\ket{\uparrow}}
  & \qw                    & \qw \\
  }} \right]
  + (1-a_\lambda)
  \left[ 
  \rule{9mm}{0mm} 
  \raisebox{11pt}{\Qcircuit @C=.5em @R=-0.3em @!R {
  & \measure{\text{click}} \\
  \lstick{ \rule[-6mm]{0mm}{7mm} \ket{\psi_1} \! \left\{ \rule{0mm}{3.5mm} \right. \!\!\!}
  & \qw                    & \qw \\
  \lstick{\ket{\downarrow}}
  & \qw                    & \qw \\
  }} \right]
  \\
  \stackrel{(\ref{a1-1})}{=}
  a_\lambda
  \left[ 
  \ket{\rule[-2mm]{0mm}{5mm} \psi_0\!} \; 
  \raisebox{8pt}{\Qcircuit @C=.5em @R=0em @!R {
  & \measure{\text{click}} \\
  & \qw                    & \qw \\
  }} \right]
  + (1-a_\lambda)
  \left[ 
  \ket{\rule[-2mm]{0mm}{5mm} \psi_1\!} \; 
  \raisebox{8pt}{\Qcircuit @C=.5em @R=0em @!R {
  & \measure{\text{click}} \\
  & \qw                    & \qw \\
  }} \right]
  \\
  =
  a_\lambda     \, \mathcal{P}_{\text{click}}\bigl(\ket{\psi_0}\bigr) +
  (1-a_\lambda) \, \mathcal{P}_{\text{click}}\bigl(\ket{\psi_1}\bigr) \,,
\end{multline}
where
\begin{equation}
\label{pr2-l1-alambda}
  a_\lambda 
  =
  \left[ 
  \rule{9mm}{0mm} \rule{0mm}{8mm}
  \raisebox{16pt}{\Qcircuit @C=.5em @R=-0.4em @!R {
  \lstick{\ket{\uparrow}}
  & \qw                   & \qw \\
  \lstick{\ket{\uparrow}\,}
  & \qw                   & \qw \\
  \lstick{ \rule{0mm}{7mm} \ket{S_\lambda} \! \left\{ \rule{0mm}{3.5mm} \right. \!\!\!}
  & \qw                   & \qw \\
  & \measureD{\uparrow} \\
  }} \right]
  =
  \left[ 
  \ket{\rule[-2mm]{0mm}{5mm} S_\lambda} 
  \raisebox{9.5pt}{\Qcircuit @C=.5em @R=0em @!R {
  & \qw                 & \qw \\
  & \measureD{\uparrow} \\
  }} \right]
  .
\end{equation}
Now we substitute Eqs.~(\ref{pr2-l1-proof1}), (\ref{pr2-l1-proof2}), 
and (\ref{pr2-l1-proof3}) into Eq.~(\ref{pr2-l1-proof4}), and get 
\begin{equation}
\label{pr2-l1-proof5}
  F_{\text{click}}\bigl((1-\lambda)\mathbf{p}_0+\lambda\mathbf{p}_1\bigr) = 
  a_\lambda\,   F_{\text{click}}\bigl(\mathbf{p}_0\bigr) +
  (1-a_\lambda) F_{\text{click}}\bigl(\mathbf{p}_1\bigr) .
\end{equation}
Since $0\leq a_\lambda \leq 1$, then the right hand side of Eq.~(\ref{pr2-l1-proof5}) 
lies between $F_{\text{click}}\bigl(\mathbf{p}_0\bigr)$ and 
$F_{\text{click}}\bigl(\mathbf{p}_1\bigr)$, that proofs the statement of Lemma~1.

\ 

{\bf Proof of Lemma~2.} Repeating all the calculations of the proof of Lemma~1 
for $\lambda=1/2$, one can get as a particular case of Eq.~(\ref{pr2-l1-proof5}) that
\begin{equation}
\label{pr2-l2-proof1}
  F_{\text{click}}\left( \frac{ \mathbf{p}_0+\mathbf{p}_1 }{2} \right) = 
  a_{1/2}\,   F_{\text{click}}\bigl(\mathbf{p}_0\bigr) +
  (1-a_{1/2}) F_{\text{click}}\bigl(\mathbf{p}_1\bigr) ,
\end{equation}
where
\begin{equation}
\label{pr2-l2-a12}
  a_{1/2} 
  =
  \left[ 
  \left(\frac{ \ket{\uparrow\uparrow}+\ket{\downarrow\downarrow} }{\sqrt{2}}\right) 
  \raisebox{9.5pt}{\Qcircuit @C=.5em @R=0em @!R {
  & \qw                 & \qw \\
  & \measureD{\uparrow} \\
  }} \right]
  .
\end{equation}
Note that the quantity $a_{1/2}$ is independent of $\mathbf{p}_0$ and $\mathbf{p}_1$. 
Therefore, Eq.~(\ref{pr2-l2-proof1}) remains valid (with the same value of $a_{1/2}$) 
if we swap $\mathbf{p}_0$ and $\mathbf{p}_1$:
\begin{equation}
\label{pr2-l2-proof1-swap}
  F_{\text{click}}\left( \frac{ \mathbf{p}_1+\mathbf{p}_0 }{2} \right) = 
  a_{1/2}\,   F_{\text{click}}\bigl(\mathbf{p}_1\bigr) +
  (1-a_{1/2}) F_{\text{click}}\bigl(\mathbf{p}_0\bigr) .
\end{equation}
Adding Eqs.~(\ref{pr2-l2-proof1}) and (\ref{pr2-l2-proof1-swap}), one can obtain
\begin{equation*}
  2\, F_{\text{click}}\left( \frac{ \mathbf{p}_0+\mathbf{p}_1 }{2} \right) = 
  F_{\text{click}}\bigl(\mathbf{p}_0\bigr) +
  F_{\text{click}}\bigl(\mathbf{p}_1\bigr) ,
\end{equation*}
Q. E. D.

\ 

{\bf Remark.} It follows from Eqs.~(\ref{pr2-l2-proof1}) and (\ref{pr2-l2-proof1-swap}) 
that $a_{1/2}=1/2$, i.~e. 
\begin{equation}
\label{pr2-bell-answer}
  \left[ 
  \left(\frac{ \ket{\uparrow\uparrow}+\ket{\downarrow\downarrow} }{\sqrt{2}}\right) 
  \raisebox{9.5pt}{\Qcircuit @C=.5em @R=0em @!R {
  & \qw                 & \qw \\
  & \measureD{\uparrow} \\
  }} \right]
  = 
  \frac12 \, .
\end{equation}

\ 

{\bf Proof of Lemma~3.} If 
$F_{\text{click}}\bigl(\mathbf{p}_0\bigr) = F_{\text{click}}\bigl(\mathbf{p}_1\bigr)$ 
then, according to Lemma~1, 
\begin{equation*}
  F_{\text{click}}\bigl((1-\lambda)\mathbf{p}_0+\lambda\mathbf{p}_1\bigr) = 
  F_{\text{click}}\bigl(\mathbf{p}_0\bigr) =
  F_{\text{click}}\bigl(\mathbf{p}_1\bigr) ,
\end{equation*}
that proves Lemma~3 for this case.

Consider the opposite case, 
$F_{\text{click}}\bigl(\mathbf{p}_0\bigr) \neq F_{\text{click}}\bigl(\mathbf{p}_1\bigr)$. 
Let us introduce the notations $\mathbf{p}_x$ and $f(x)$:
\begin{gather}
  \mathbf{p}_x = (1-x)\,\mathbf{p}_0+x\,\mathbf{p}_1, \\
  f(x) = 
  \frac {
    F_{\text{click}}\bigl(\mathbf{p}_x\bigr) - F_{\text{click}}\bigl(\mathbf{p}_0\bigr)
  }{
    F_{\text{click}}\bigl(\mathbf{p}_1\bigr) - F_{\text{click}}\bigl(\mathbf{p}_0\bigr)    
  } \, ,
\end{gather}
where $x$ is any number in the range $[0,1]$. By definitions, 
\begin{equation}
\label{pr2-l3-proof1}
  f(0)=0, \quad f(1)=1.
\end{equation}
Let us consider two numbers $x_1\in[0,1]$ and $x_2\in[0,1]$. Applying Lemma~2 to the 
vectors $\mathbf{p}_{x_1}$ and $\mathbf{p}_{x_2}$, one can see that
\begin{equation}
\label{pr2-l3-proof2}
  f\left( \frac{x_1+x_2}{2} \right) = \frac{f(x_1)+f(x_2)}{2} .
\end{equation}
Then, one can apply Lemma~1 to the vectors $\mathbf{p}_0$ and $\mathbf{p}_1$ taking 
$\lambda=x_2$. This gives the inequality 
\begin{equation*}
  0 \leq f(x_2) \leq 1 .
\end{equation*}
If $x_1 \leq x_2$, one can apply Lemma~1 to vectors $\mathbf{p}_0$ and $\mathbf{p}_{x_2}$ 
taking $\lambda=x_1/x_2$, with the following result:
\begin{equation}
\label{pr2-l3-proof3}
  0 \leq x_1 \leq x_2 \leq 1   \quad\Rightarrow\quad   0 \leq f(x_1) \leq f(x_2) \leq 1.
\end{equation}
Therefore the function $f(x)$ is monotonically increasing in the range $x\in[0,1]$.

Let us prove that $f(x) \equiv x$ on the basis of Eqs.~(\ref{pr2-l3-proof1}), 
(\ref{pr2-l3-proof2}), (\ref{pr2-l3-proof3}). First, we apply Eq.~(\ref{pr2-l3-proof2}) 
with $x_1=0$, $x_2=1$:
\[
  f\left(\frac12\right) = \frac{f(0)+f(1)}{2} = \frac12.
\]
Then, Eq.~(\ref{pr2-l3-proof2}) can be applied once again with $x_1=0$, $x_2=1/2$, 
and with $x_1=1/2$, $x_2=1$:
\begin{gather*}
  f\left(\frac14\right) = \frac{f(0)+f(1/2)}{2} = \frac14, \\
  f\left(\frac34\right) = \frac{f(1/2)+f(1)}{2} = \frac34,
\end{gather*}
and so on. As a result, we get the equality
\begin{equation}
\label{pr2-l3-proof4}
  f\left( \frac{p}{2^k} \right) = \frac{p}{2^k}
\end{equation}
for any $k=0,1,2,\ldots,$ and any $p=0,1,\ldots,2^k$. 

Now let us consider some value $x\in[0,1)$ and some natural number $k$, and introduce two 
quantities 
\begin{equation}
\label{pr2-l3-proof5}
  x_- = \frac{[2^kx]}{2^k} \,, \quad x_+ = \frac{[2^kx]+1}{2^k} \,,
\end{equation}
where the square brackets denote the integer part. According to Eq.~(\ref{pr2-l3-proof4}),
\begin{equation}
\label{pr2-l3-proof6}
  f(x_-) = x_-, \quad f(x_+) = x_+ \,.
\end{equation}
Since
\begin{equation}
\label{pr2-l3-proof7}
  0 \leq x_- \leq x \leq x_+ \leq 1,
\end{equation}
and $f(x)$ is a monotonically increasing function, then
\begin{equation}
\label{pr2-l3-proof8}
  f(x_-) \leq f(x) \leq f(x_+) ,
\end{equation}
or, taking Eq.~(\ref{pr2-l3-proof6}) into account,
\begin{equation}
\label{pr2-l3-proof9}
  x_- \leq f(x) \leq x_+ \,.
\end{equation}
One can see from Eqs.~(\ref{pr2-l3-proof7}) and (\ref{pr2-l3-proof9}) that both $x$ 
and $f(x)$ are bound within the range $[x_-,x_+]$. Therefore
\begin{equation}
\label{pr2-l3-proof10}
  |f(x)-x| \leq x_+-x_- = 2^{-k}.
\end{equation}
As the number $k$ can be arbitrarily large, Eq.~(\ref{pr2-l3-proof10}) actually means that 
\[
  |f(x)-x| = 0.
\]
So $f(x)=x$ for any $x\in[0,1)$. This equality is valid also for $x=1$, due to 
Eq.~(\ref{pr2-l3-proof1}). Therefore $f(\lambda)=\lambda$, i.~e.
\[
  \frac {
    F_{\text{click}}\bigl( (1-\lambda)\,\mathbf{p}_0+\lambda\,\mathbf{p}_1 \bigr) - 
    F_{\text{click}}\bigl(\mathbf{p}_0\bigr)
  }{
    F_{\text{click}}\bigl(\mathbf{p}_1\bigr) - F_{\text{click}}\bigl(\mathbf{p}_0\bigr)    
  } = \lambda ,
\]
i.~e.
\[
  F_{\text{click}}\bigl((1-\lambda)\mathbf{p}_0+\lambda\mathbf{p}_1\bigr) = 
  (1-\lambda)F_{\text{click}}(\mathbf{p}_0) + \lambda\, F_{\text{click}}(\mathbf{p}_1) ,
\]
Q.E.D.

\ 

\begin{figure}
\includegraphics[width=2in]{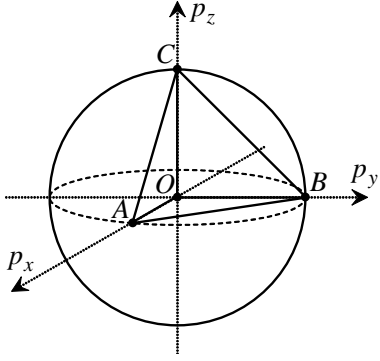}
\caption{Reference points $O,A,B,C$ in the Bloch ball.}
\label{fig:ball}
\end{figure}

Using Lemma~3, it is easy to prove that the function $F_{\text{click}}(\mathbf{p})$ 
is \emph{linear}. For this, we choose four reference points $O,A,B,C$ in the Bloch ball 
(see Fig.~\ref{fig:ball})
corresponding to the following vectors of spin polarization:
\begin{gather*}
  \mathbf{p}_O = (0,0,0), \\
  \mathbf{p}_A = (1,0,0), \\
  \mathbf{p}_B = (0,1,0), \\
  \mathbf{p}_C = (0,0,1),
\end{gather*}
and introduce four parameters 
\begin{align*}
  \alpha_x &= F_{\text{click}}(\mathbf{p}_A) - F_{\text{click}}(\mathbf{p}_O), \\
  \alpha_y &= F_{\text{click}}(\mathbf{p}_B) - F_{\text{click}}(\mathbf{p}_O), \\
  \alpha_z &= F_{\text{click}}(\mathbf{p}_C) - F_{\text{click}}(\mathbf{p}_O), \\
  \beta    &= F_{\text{click}}(\mathbf{p}_O).
\end{align*}

Let us show that
the probability $F_{\text{click}}(\mathbf{p})$ for any $\mathbf{p}$ 
can be expressed through the parameters 
$\alpha_x,\alpha_y,\alpha_z,\beta$. This 
can be done in four steps:

\noindent (i) for any $\mathbf{p}=(p_x,0,0)$ in the line segment $OA$ one can find that 
\[
  F_{\text{click}}(\mathbf{p}) = \alpha_x p_x + \beta
\]
by applying Lemma~3 with parameters 
\[
  \mathbf{p}_0=(0,0,0), \; \mathbf{p}_1=(1,0,0), \;
  \lambda=p_x;
\]

\noindent (ii) for any $\mathbf{p}=(p_x,p_y,0)$ in the triangle $OAB$ one can find that 
\[
  F_{\text{click}}(\mathbf{p}) = \alpha_x p_x + \alpha_y p_y + \beta
\]
by applying Lemma~3 with parameters 
\[
  \mathbf{p}_0=\frac{(p_x,0,0)}{1-p_y}, \; \mathbf{p}_1=(0,1,0), \;
  \lambda=p_y;
\]

\noindent (iii) for any $\mathbf{p}=(p_x,p_y,p_z)$ in the tetrahedron $OABC$ one can find that 
\begin{equation}
\label{pr2-fclick-linear}
  F_{\text{click}}(\mathbf{p}) = \alpha_x p_x + \alpha_y p_y + \alpha_z p_z + \beta 
  \equiv \boldsymbol{\alpha} \cdot \mathbf{p} + \beta
\end{equation}
by applying Lemma~3 with parameters 
\[
  \mathbf{p}_0=\frac{(p_x,p_y,0)}{1-p_z}, \; \mathbf{p}_1=(0,0,1), \;
  \lambda=p_z;
\]

\noindent (iv) finally, Eq.~(\ref{pr2-fclick-linear}) can be estabilshed for an arbitrary point 
$\mathbf{p}$ in the Bloch ball by drawing a line that passes through the point $\mathbf{p}$ 
and intersects the tetrahedron $OABC$, choosing two different points $\mathbf{q}_1$ and $\mathbf{q}_2$ 
lying on this line and belonging the tetrahedron, and applying Lemma~3 with parameters 
\[
  \mathbf{p}_0=\mathbf{p}, \; \mathbf{p}_1=\mathbf{q}_2, \;
  \lambda=\frac{|\mathbf{p}-\mathbf{q}_1|}{|\mathbf{p}-\mathbf{q}_2|} \,.
\]

Eq.~(\ref{pr2-fclick-linear}) proves that the function $F_{\text{click}}(\mathbf{p})$ is linear. 
Substituting it into Eq.~(\ref{pr2-function1}), one can get the most general expression for the probability 
$\mathcal{P}_{\text{click}}$ of the ``click'' event, as a function of the state vector 
$\ket{\mathbf{\Psi}}$:
\begin{equation}
\label{pr2-linear-pure}
  \mathcal{P}_{\text{click}}\bigl(\ket{\mathbf{\Psi}}\bigr) = 
  \boldsymbol{\alpha} \cdot \mathbf{p}\bigl(\ket{\mathbf{\Psi}}\bigr) + \beta,
\end{equation}
where a vector $\boldsymbol{\alpha}=(\alpha_x,\alpha_y,\alpha_z)$ and a number $\beta$ 
are characteristics of the measuring device.

We considered so far only \emph{pure} states of the combined system ``spin + environment''. 
Now we will generalize the results to the case of \emph{mixed} states. Let the state 
$\mathcal{S}$ be a statistical mix of pure states $\ket{\mathbf{\Psi}_k}$ taken 
with probabilities $P_k$. The polarization vector $\mathbf{p}\bigl(\mathcal{S}\bigr)$ 
in this state is a weighted average of spin polarizations in the states $\ket{\mathbf{\Psi}_k}$:
\begin{equation}
\label{pr2-mixed-polarization}
  \mathbf{p}\bigl(\mathcal{S}\bigr) = \sum_k P_k\, \mathbf{p}\bigl(\ket{\mathbf{\Psi}_k}\bigr).
\end{equation}
The probability $\mathcal{P}_{\text{click}}\bigl(\mathcal{S}\bigr)$ of the ``click'' event 
in the state $\mathcal{S}$ can be found by the law of total probability:
\begin{equation}
\label{pr2-mixed-pclick}
  \mathcal{P}_{\text{click}}\bigl(\mathcal{S}\bigr) = 
  \sum_k P_k\, \mathcal{P}_{\text{click}}\bigl(\ket{\mathbf{\Psi}_k}\bigr).
\end{equation}
Substitution of Eq.~(\ref{pr2-linear-pure}) into Eq.~(\ref{pr2-mixed-pclick}) gives
\begin{equation}
\label{pr2-mixed-pclick1}
  \mathcal{P}_{\text{click}}\bigl(\mathcal{S}\bigr) = 
  \boldsymbol{\alpha} \cdot 
  \left( \sum_k P_k\, \mathbf{p}\bigl(\ket{\mathbf{\Psi}_k}\bigr) \right) 
  + \beta,
\end{equation}
that is,
\begin{equation}
\label{pr2-mixed-pclick2}
  \mathcal{P}_{\text{click}}\bigl(\mathcal{S}\bigr) = 
  \boldsymbol{\alpha} \cdot \mathbf{p}\bigl(\mathcal{S}\bigr) + \beta.
\end{equation}

So the probability $\mathcal{P}_{\text{click}}$ is a linear function of the 
spin polarization, not only for pure states, but for mixed states as well.

It was mentioned in the beginning of this Subsection, that any outcome of 
any measuring device (interacting with the spin) can play the role of the 
``click''. Therefore the results of this Subsection are valid for any 
outcome. We summarize them in the following Theorem.

\ 

{\bf Theorem 1.} It follows from the five Assumptions listed in Subsection~\ref{sub:st3-assump} that, 
for any measuring device interacting to the spin degree of freedom of 
a spin-1/2 particle, the probability of any measurement result is a linear function 
of the spin polarization vector of this particle.

\ 

This Theorem is the central result of the present paper. In the next two Subsections, we will 
use it for proving the Born rule for the spin projection measurement, as well as for 
the measurement of the particle coordinate.

Theorem~1 can be formulated also in the language of the reduced density matrix $\hat\rho$ 
of the measured spin, which is connected to the spin polarization $\mathbf{p}$ as follows:
\begin{equation}
\label{pr2-dmatrix}
  \hat\rho = \frac{1+\hat{\boldsymbol{\sigma}}\cdot\mathbf{p}}{2} \equiv \frac{1}{2} 
  \left( \begin{array}{cc} 1+p_z & p_x-ip_y \\ p_x+ip_y & 1-p_z \end{array} \right).
\end{equation}
Namely, Theorem~1 states that the probability is a linear function of the real and imaginary 
parts of the density matrix $\hat\rho$.
It is easy to find an explicit form for this linear function. One can see that 
Eq.~(\ref{pr2-mixed-pclick2}) can be presented in the following equivalent form:
\begin{equation}
\label{pr2-povm}
  \mathcal{P}_{\text{click}}(\mathcal{S}) = 
  \text{Tr} \left( \hat M_{\text{click}} \; \hat\rho(\mathcal{S}) \right),
\end{equation}
where $\hat M_{\text{click}}$ is a non-negative self-adjoint operator:
\begin{equation}
\label{pr2-m}
  \hat M_{\text{click}} = 
  \boldsymbol{\alpha} \cdot \hat{\boldsymbol{\sigma}} + \beta \equiv 
  \left( \begin{array}{cc} \beta+\alpha_z & \alpha_x-i\alpha_y \\ 
  \alpha_x+i\alpha_y & \beta-\alpha_z \end{array} \right),
\end{equation}
and the density matrix $\hat\rho(\mathcal{S})$ is given by Eq.~(\ref{pr2-dmatrix}).

\subsection{Proof of the Born rule for spin-1/2 projection measurement}
\label{sub:pr3-spin}

\begin{figure}
\includegraphics[width=2in]{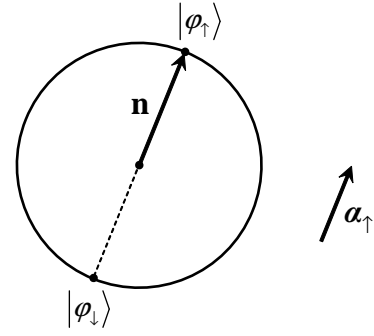}
\caption{The points when the probability $\mathcal{P}_\uparrow$ 
reaches its maximal value $\mathcal{P}_{\uparrow\text{max}}$ 
($\mathbf{p}=\mathbf{n}$, the state vector $\ket{\varphi_\uparrow}$) 
and its minimal value $\mathcal{P}_{\uparrow\text{min}}$ 
($\mathbf{p}=-\mathbf{n}$, the state vector $\ket{\varphi_\downarrow}$)
in the Bloch ball. The vector $\mathbf{n}$ is the unit vector in the direction of 
$\boldsymbol{\alpha}_\uparrow$.}
\label{fig:ball2}
\end{figure}

Now we consider measurement of a spin-1/2 particle by some device having two possible 
outcomes: $\uparrow$ and $\downarrow\,.$ According to Theorem~1, the probabilities 
$\mathcal{P}_\uparrow$ and $\mathcal{P}_\downarrow$ 
of these outcomes are linear functions of the spin polarization $\mathbf{p}$. 
In particular,
\begin{equation}
\label{pr3-p1}
  \mathcal{P}_\uparrow = \boldsymbol{\alpha}_\uparrow \cdot \mathbf{p} + \beta_\uparrow ,
\end{equation}
where the vector $\boldsymbol{\alpha}_\uparrow$ and the scalar $\beta_\uparrow$ are 
characteristics of the measuring device. Let us denote as $\mathbf{n}$ the unit vector directed along 
$\boldsymbol{\alpha}_\uparrow$:
\begin{equation}
\label{pr3-n}
  \mathbf{n} = \boldsymbol{\alpha}_\uparrow / |\boldsymbol{\alpha}_\uparrow| .
\end{equation}
As the polarization vector $\mathbf{p}$ is bound in the Bloch ball 
$(|\mathbf{p}|\leq1)$, then the maximal value $\mathcal{P}_{\uparrow\text{max}}$ of 
the probability $\mathcal{P}_\uparrow$ is reached when the vector $\mathbf{p}$ is directed 
along $\boldsymbol{\alpha}_\uparrow$ and has its maximal length $|\mathbf{p}|=1$, i.~e. 
when $\mathbf{p}=\mathbf{n}$. Therefore,
\begin{equation}
\label{pr3-pmax}
  \mathcal{P}_{\uparrow\text{max}} = 
  \boldsymbol{\alpha}_\uparrow \cdot \mathbf{n} + \beta_\uparrow =
  |\boldsymbol{\alpha}_\uparrow| + \beta_\uparrow .
\end{equation}
Analogously, the minimal value $\mathcal{P}_{\uparrow\text{min}}$ of 
the probability $\mathcal{P}_\uparrow$ is reached when $\mathbf{p}=-\mathbf{n}$:
\begin{equation}
\label{pr3-pmin}
  \mathcal{P}_{\uparrow\text{min}} = 
  \boldsymbol{\alpha}_\uparrow \cdot (-\mathbf{n}) + \beta_\uparrow =
  -|\boldsymbol{\alpha}_\uparrow| + \beta_\uparrow .
\end{equation}
Since the maximal and minimal values of $\mathcal{P}_\uparrow$ are reached at the maximal spin 
polarization $(|\mathbf{p}|=1)$, then they correspond to some \emph{pure} states of the spin. 
Let us denote corresponding unit spinors as $\ket{\varphi_\uparrow}$ and $\ket{\varphi_\downarrow}$:
\begin{equation}
\label{pr3-phi1}
  \mathbf{p} \bigl( \ket{\varphi_\uparrow} \bigr) = \mathbf{n} , \quad
  \mathbf{p} \bigl( \ket{\varphi_\downarrow} \bigr) = -\mathbf{n} 
\end{equation}
(see Fig.~\ref{fig:ball2}). Remembering the definition~(\ref{pr2-p-def}) of the vector $\mathbf{p}$, 
one can rewrite Eq.~(\ref{pr3-phi1}) as
\begin{equation}
\label{pr3-phi2}
  \bra{\varphi_\uparrow} \hat{\boldsymbol{\sigma}} \ket{\varphi_\uparrow} = \mathbf{n} , \quad
  \bra{\varphi_\downarrow} \hat{\boldsymbol{\sigma}} \ket{\varphi_\downarrow} = -\mathbf{n} .
\end{equation}
Note that $\ket{\varphi_\uparrow}$ and $\ket{\varphi_\downarrow}$ are orthogonal to each other: 
\begin{equation}
\label{pr3-phi-ort}
  \langle\varphi_\uparrow|\varphi_\downarrow\rangle = 0 ,
\end{equation}
because they correspond to the opposite spin directions.

Using Eqs.~(\ref{pr3-n}), (\ref{pr3-pmax}), (\ref{pr3-pmin}), (\ref{pr3-phi2}), 
one can express parameters $\boldsymbol{\alpha}_\uparrow$ and $\beta_\uparrow$ via 
$\mathcal{P}_{\uparrow\text{max}}$, $\mathcal{P}_{\uparrow\text{min}}$, 
and $\ket{\varphi_\uparrow}$:
\begin{gather}
\label{pr3-alpha}
  \boldsymbol{\alpha}_\uparrow =  
  \frac{\mathcal{P}_{\uparrow\text{max}}-\mathcal{P}_{\uparrow\text{min}}}{2} \,
  \bra{\varphi_\uparrow} \hat{\boldsymbol{\sigma}} \ket{\varphi_\uparrow} ,
\\
\label{pr3-beta}
  \beta_\uparrow =
  \frac{\mathcal{P}_{\uparrow\text{max}}+\mathcal{P}_{\uparrow\text{min}}}{2} \,.
\end{gather}
With these expressions for $\boldsymbol{\alpha}_\uparrow$ and $\beta_\uparrow$, 
Eq.~(\ref{pr3-p1}) takes the following form:
\begin{equation}
\label{pr3-p2}
  \mathcal{P}_\uparrow = 
  \frac{\mathcal{P}_{\uparrow\text{max}}-\mathcal{P}_{\uparrow\text{min}}}{2} 
  \bra{\varphi_\uparrow} \hat{\boldsymbol{\sigma}} \ket{\varphi_\uparrow} \cdot \mathbf{p} 
  + \frac{\mathcal{P}_{\uparrow\text{max}}+\mathcal{P}_{\uparrow\text{min}}}{2} \,.
\end{equation}
Finally, the quantity 
$\bra{\varphi_\uparrow} \hat{\boldsymbol{\sigma}} \ket{\varphi_\uparrow} \cdot \mathbf{p}$ 
can be expressed via the spin density matrix $\hat\rho$ using Eq.~(\ref{pr2-dmatrix}):
\begin{equation}
\label{pr3-rho}
  \bra{\varphi_\uparrow} \hat{\boldsymbol{\sigma}} \ket{\varphi_\uparrow} \cdot \mathbf{p} = 
  2 \bra{\varphi_\uparrow} \hat\rho \ket{\varphi_\uparrow} - 1 .
\end{equation}
Substituting this into Eq.~(\ref{pr3-p2}), we get the following result:
\begin{equation}
\label{pr3-p3-up}
  \mathcal{P}_\uparrow = 
  \bigl(\mathcal{P}_{\uparrow\text{max}}-\mathcal{P}_{\uparrow\text{min}}\bigr) 
  \bra{\varphi_\uparrow} \hat\rho \ket{\varphi_\uparrow} 
  + \mathcal{P}_{\uparrow\text{min}} .
\end{equation}
Analogous expression can be obtained for the probability 
$\mathcal{P}_\downarrow\equiv1-\mathcal{P}_\uparrow$: 
\begin{equation}
\label{pr3-p3-down}
  \mathcal{P}_\downarrow = 
  \bigl(\mathcal{P}_{\downarrow\text{max}}-\mathcal{P}_{\downarrow\text{min}}\bigr) 
  \bra{\varphi_\downarrow} \hat\rho \ket{\varphi_\downarrow} 
  + \mathcal{P}_{\downarrow\text{min}} ,
\end{equation}
where
\begin{equation}
\label{pr3-p-updown}
  \mathcal{P}_{\downarrow\text{max}} = 1 - \mathcal{P}_{\uparrow\text{min}} , \quad
  \mathcal{P}_{\downarrow\text{min}} = 1 - \mathcal{P}_{\uparrow\text{max}} .
\end{equation}
If the spin is in some pure state $\ket{\psi}$, the density matrix is equal to
\begin{equation*}
  \hat\rho = \ket{\psi}\bra{\psi}
\end{equation*}
(the state vector $\ket{\psi}$ is assumed to be normalized). 
Correspondingly, Eqs.~(\ref{pr3-p3-up}) and (\ref{pr3-p3-down}) for pure states take the form
\begin{gather}
\label{pr3-p4-up}
  \mathcal{P}_\uparrow = 
  \bigl(\mathcal{P}_{\uparrow\text{max}}-\mathcal{P}_{\uparrow\text{min}}\bigr) 
  \left| \langle\varphi_\uparrow|\psi\rangle \right|^2 
  + \mathcal{P}_{\uparrow\text{min}} ,
\\
\label{pr3-p4-down}
  \mathcal{P}_\downarrow = 
  \bigl(\mathcal{P}_{\downarrow\text{max}}-\mathcal{P}_{\downarrow\text{min}}\bigr) 
  \left| \langle\varphi_\downarrow|\psi\rangle \right|^2 
  + \mathcal{P}_{\downarrow\text{min}} .
\end{gather}
It is clearly seen from Eqs.~(\ref{pr3-p3-up})--(\ref{pr3-p4-down}) that if (and only if)
\begin{equation}
  \mathcal{P}_{\uparrow\text{max}} = 1 \quad \text{and} \quad \mathcal{P}_{\uparrow\text{min}} = 0 ,
\end{equation}
then the probabilities $\mathcal{P}_\uparrow$ and $\mathcal{P}_\downarrow$ obey the Born rule---namely, 
probabilities for any state $\mathcal{S}$ characterized by the spin density matrix $\hat\rho$ are
\begin{equation}
\label{pr3-bornrule-mixed}
  \mathcal{P}_\uparrow(\mathcal{S}) = 
  \bra{\varphi_\uparrow} \hat\rho \ket{\varphi_\uparrow} , \quad
  \mathcal{P}_\downarrow(\mathcal{S}) = 
  \bra{\varphi_\downarrow} \hat\rho \ket{\varphi_\downarrow} ,
\end{equation}
and probabilities for any pure state $\ket{\psi}$ of the spin are
\begin{equation}
\label{pr3-bornrule-pure}
  \mathcal{P}_\uparrow \bigl( \ket{\psi} \bigr) = 
  \left| \langle\varphi_\uparrow|\psi\rangle \right|^2 \,, \quad 
  \mathcal{P}_\downarrow \bigl( \ket{\psi} \bigr) = 
  \left| \langle\varphi_\downarrow|\psi\rangle \right|^2 \,.
\end{equation}

So the nesessary and sufficient condition for the Born rule is the existence of such two 
\emph{eigenvectors} $\ket{\varphi_\uparrow}$ and $\ket{\varphi_\downarrow}$, that 
\begin{equation}
  \mathcal{P}_\uparrow \bigl( \ket{\varphi_\uparrow} \bigr) = 1 \quad \text{and} \quad
  \mathcal{P}_\downarrow \bigl( \ket{\varphi_\downarrow} \bigr) = 1,
\end{equation}
i.~e. that the measurement results are \emph{predictable} for these states: 
$\uparrow$ for $\ket{\varphi_\uparrow}$, 
and $\downarrow$ for $\ket{\varphi_\downarrow}$ with probability 1. 
According to Eq.~(\ref{pr3-phi-ort}), 
the eigenvectors are necessarily orthogonal to each other. One can thus say 
that the two eigenvectors define the spin projection which is measured by the 
given device.

The Stern--Gerlach apparatus is an example of a device that measures the 
vertical projection of the spin. As was mentioned in Subsection~\ref{sub:st1-stern}, 
if the spin is in the pure state 
$\ket{\uparrow}$ before the measurement, then the measurement result 
should be nesessary ``spin up''. Analogously, the measurement of the spin in the state 
$\ket{\downarrow}$ gives the result ``spin down'' 
with probability 1. Hence the vectors $\ket{\uparrow}$ and $\ket{\downarrow}$ 
are eigenvectors for the Stern--Gerlach apparatus: 
$\mathcal{P}_\uparrow \bigl(\ket{\uparrow}\bigr) = 
\mathcal{P}_\downarrow \bigl(\ket{\downarrow}\bigr) = 1$.

The results of this Subsection are summarized in the following Theorem:

\ 

{\bf Theorem 2.} {\bf If} three conditions (i)--(iii) are satisfied:
\newline (i) five Assumptions of Subsection~\ref{sub:st3-assump} are fulfilled,
\newline (ii) an apparatus $\mathcal{A}$ interacts with the spin degree of 
freedom of a spin-1/2 particle and has two possible classical outcomes 
$\uparrow$ and $\downarrow\,$,
\newline (iii) there are such two states $\mathcal{S}_\uparrow$ and 
$\mathcal{S}_\downarrow$ of the spin, that the results of measurements by the 
apparatus $\mathcal{A}$ are \emph{predictable} for these states: 
$\uparrow$ for $\mathcal{S}_\uparrow$ 
and $\downarrow$ for $\mathcal{S}_\downarrow$ with probability 1,
\newline {\bf then}
\newline (a) $\mathcal{S}_\uparrow$ and $\mathcal{S}_\downarrow$ are \emph{pure} 
states of the spin, and they can be charactrized by normalized state vectors 
$\ket{\varphi_\uparrow}$ and $\ket{\varphi_\downarrow}$ correspondingly,
\newline (b) the vectors $\ket{\varphi_\uparrow}$ and $\ket{\varphi_\downarrow}$ 
are orthogonal to each other, 
\newline (c) when the spin is measured by the apparatus $\mathcal{A}$, 
the probabilities $\mathcal{P}_\uparrow$ and $\mathcal{P}_\downarrow$ of the 
outcomes $\uparrow$ and $\downarrow$ obey the Born rule~(\ref{pr3-bornrule-pure}) 
for any pure state $\ket{\psi}$ of the spin, and obey its 
generalization~(\ref{pr3-bornrule-mixed}) for any mixed state.

\ 

Based on Theorem~2, we can answer the Questions asked in the beginning of Section~\ref{sec:st}:

{\bf Answer to Question~1.} For different devices $\mathcal{A}_1$ and $\mathcal{A}_2$ that measure 
the same spin projection (i.~e. there are such two states $\ket{\varphi_\uparrow}$ and 
$\ket{\varphi_\downarrow}$ that 
$\mathcal{P}_\uparrow(\ket{\varphi_\uparrow},\mathcal{A}_1)=
\mathcal{P}_\uparrow(\ket{\varphi_\uparrow},\mathcal{A}_2)=1$ and 
$\mathcal{P}_\downarrow(\ket{\varphi_\downarrow},\mathcal{A}_1)=
\mathcal{P}_\downarrow(\ket{\varphi_\downarrow},\mathcal{A}_2)=1$),
probabilities are the same for any state $\mathcal{S}$:
\begin{equation*}
  \forall\,\mathcal{S} \;\;
  \mathcal{P}_\uparrow(\mathcal{S},\mathcal{A}_1) =
  \mathcal{P}_\uparrow(\mathcal{S},\mathcal{A}_2) , \;\;
  \mathcal{P}_\downarrow(\mathcal{S},\mathcal{A}_1) =
  \mathcal{P}_\downarrow(\mathcal{S},\mathcal{A}_2) .
\end{equation*}

{\bf Answer to Question~2.} If an apparatus $\mathcal{A}$ obeys the conditions of Theorem~2, 
then it measures some projection of the spin-1/2, and 
the probabilities $\mathcal{P}_\uparrow(\mathcal{S},\mathcal{A})$ and 
$\mathcal{P}_\downarrow(\mathcal{S},\mathcal{A})$ are given by 
Eq.~(\ref{pr3-bornrule-mixed}) for any state $\mathcal{S}$, and by (\ref{pr3-bornrule-pure}) 
for pure states. 
In a more general case of \emph{arbitrary} apparatus interacting with the spin 
degree of freedom (for spin 1/2), the probability of any measurement result is a 
linear function of the spin polarization.

\subsection{Proof of the Born rule for coordinate measurement} 
\label{sub:pr4-coord}

We considered so far only measurements of the spin degree of freedom, and only in the 
case of spin 1/2. Now we will show that the obtained results can be generalized to 
measurements of other systems.

Let $\{\ket{\varphi_0}, \ket{\varphi_1}, \ket{\varphi_2}, \ldots\}$ be an orthonormal 
basis in the state space of some quantum object $\mathcal{O}$. Let there be a 
measuring device that, being applied to $\mathcal{O}$, always ``clicks'' when the 
object $\mathcal{O}$ is in the state $\ket{\varphi_1}$, and never ``clicks'' when the 
object $\mathcal{O}$ is in the state $\ket{\varphi_0}$:
\begin{equation}
  \mathcal{P}_{\text{click}}\bigl(\ket{\varphi_0}\bigr) = 0, \quad
  \mathcal{P}_{\text{click}}\bigl(\ket{\varphi_1}\bigr) = 1,
\end{equation}
where $\mathcal{P}_{\text{click}}$ denotes the probability of ``clicking''. The question 
is: what is the probability of ``clicking'' when the quantum object is in some superposition 
of states $\ket{\varphi_0}$ and $\ket{\varphi_1}$:
\begin{equation}
  \mathcal{P}_{\text{click}}\bigl(c_0\ket{\varphi_0}+c_1\ket{\varphi_1}\bigr) = \; ?
\end{equation}
Here $c_0$ and $c_1$ are complex numbers, $|c_0|^2+|c_1|^2=1$.

As in Subsection~\ref{sub:pr2-polariz}, one needs consideration of a larger system to answer 
this question. Let an ``environment'' $\mathcal{E}$ be a large enough quanum system containing 
at least three spin-1/2 particles. We will consider a composite system $\mathcal{O}+\mathcal{E}$. 
Each state vector of the composite system can be represented as 
\begin{equation}
  \ket{\varphi_0}\ket{\boldsymbol{\chi}_0} + \ket{\varphi_1}\ket{\boldsymbol{\chi}_1} + 
  \ket{\varphi_2}\ket{\boldsymbol{\chi}_2} + \ldots ,
\end{equation}
where $\ket{\boldsymbol{\chi}_0}$ etc. are state vectors of $\mathcal{E}$. Among all state 
vectors of the system $\mathcal{O}+\mathcal{E}$, we select the subspace $\tilde{\mathbb{S}}$ of 
vectors containing only contributions of $\ket{\varphi_0}$ and $\ket{\varphi_1}$, i.e. 
having the form
\begin{equation}
\label{pr4-s-tilde}
  \ket{\varphi_0}\ket{\boldsymbol{\chi}_0}+\ket{\varphi_1}\ket{\boldsymbol{\chi}_1}
\end{equation}
with arbitrary state vectors $\ket{\boldsymbol{\chi}_0},\ket{\boldsymbol{\chi}_1}$ of the 
``environment''. One can see that the subspace $\tilde{\mathbb{S}}$ 
is formally equivalent to the space $\mathbb{S}$ of state vectors 
of the system ``spin-1/2 + environment''. Indeed, each vector of $\mathbb{S}$ can be 
represented as
\begin{equation}
\label{pr4-s}
  \ket{\downarrow}\ket{\boldsymbol{\chi}_0}+\ket{\uparrow}\ket{\boldsymbol{\chi}_1} ,
\end{equation}
and it is clearly seen from Eqs.~(\ref{pr4-s-tilde}) and (\ref{pr4-s}) that there is 
one-to-one correspondence between vectors of $\tilde{\mathbb{S}}$ and vectors of $\mathbb{S}$.

As a result of the similarity between vector spaces $\mathbb{S}$ and $\tilde{\mathbb{S}}$, 
one can introduce an ``isospin'' $\tilde{\mathbf{s}}$ (an observable having all formal properties 
of the spin 1/2) on $\tilde{\mathbb{S}}$. Namely, the vectors 
$\ket{\varphi_0}\ket{\boldsymbol{\chi}}$ and $\ket{\varphi_1}\ket{\boldsymbol{\chi}}$ (with 
any $\ket{\boldsymbol{\chi}}$) can be considered as eigenvectors of $\tilde s_z$ with the 
corresponding eigenvalues $-1/2$ and $+1/2$. Then, one can define the isospin ``polarization vector'' 
$\tilde{\mathbf{p}}$ for any state (\ref{pr4-s-tilde}) of $\tilde{\mathbb{S}}$, as a quantity 
equal to the spin polarization $\mathbf{p}$ for the corresponding state (\ref{pr4-s}) of $\mathbb{S}$. 
More explicitly, the components $(\tilde p_x,\tilde p_y,\tilde p_z)$ of $\tilde{\mathbf{p}}$ are
\begin{align*}
  \tilde p_x &= 2\,\text{Re}\, \langle \boldsymbol{\chi}_1|\boldsymbol{\chi}_0 \rangle , \\
  \tilde p_y &= 2\,\text{Im}\, \langle \boldsymbol{\chi}_1|\boldsymbol{\chi}_0 \rangle , \\
  \tilde p_z &= \langle \boldsymbol{\chi}_1|\boldsymbol{\chi}_1 \rangle - 
  \langle \boldsymbol{\chi}_0|\boldsymbol{\chi}_0 \rangle .
\end{align*}

Now, all the derivations of Subsection~\ref{sub:pr2-polariz} can be literally repeated for the 
system $\mathcal{O}+\mathcal{E}$ (with the restricted vector state $\tilde{\mathbb{S}}$), 
replacing the spin polarization 
$\mathbf{p}$ with the ``isospin polarization'' $\tilde{\mathbf{p}}$. 
As a result, one can see that the probability $\mathcal{P}_{\text{click}}$ 
is a linear function of $\tilde{\mathbf{p}}$ for all state vectors of $\tilde{\mathbb{S}}$. 
Then, all considerations of Subsection~\ref{sub:pr3-spin} can be applied to the state vectors 
of $\tilde{\mathbb{S}}$ (with 
$\mathcal{P}_{\text{click}}$, $\tilde{\mathbf{p}}$, $\ket{\varphi_1}$ and $\ket{\varphi_0}$ 
instead of 
$\mathcal{P}_\uparrow$, $\mathbf{p}$, $\ket{\varphi_\uparrow}$ and $\ket{\varphi_\downarrow}$). 
So, for a pure state 
\begin{equation}
  \ket{\psi} = c_0\ket{\varphi_0}+c_1\ket{\varphi_1} 
\end{equation}
of the system $\mathcal{O}$, we obtain the result
\begin{equation}
\label{pr4-pclick-answer}
  \mathcal{P}_{\text{click}}\bigl(\ket{\psi}\bigr) = 
  |\langle\varphi_1|\psi\rangle|^2 = 
  |c_1|^2 \,,
\end{equation}
as an analog of Eq.~(\ref{pr3-bornrule-pure}). 

Let is illustrate this idea on the example of measurement of the particle position. 
We suppose for simplicity that the particle has no internal degrees of freedom, and 
its wavefunction $\psi$ is one-dimensional: $\psi=\psi(x)$. As a measuring device, we 
consider an ideal detector that catches everything within the range $[x_1,x_2]$ of 
the $x$-coordinate, and beeps when it catches a particle. 
So the probability of the beep can be interpreted as a probability of finding a particle in the 
coordinate range $[x_1,x_2]$. We denote this probability (for the wavefunction $\psi(x)$) 
as $\mathcal{P}[\psi]$.

We postulate two obvious properties of the detector:
\newline {\bf1.} If the wavefunction of the particle vanishes for all $x\in[x_1,x_2]$, then 
the detector \emph{never} beeps ($\mathcal{P}=0$).
\newline {\bf2.} If the wavefunction vanishes for all $x$ \emph{outside} the range $[x_1,x_2]$, then 
the detector \emph{always} beeps ($\mathcal{P}=1$).

\begin{figure}
\includegraphics[width=3in]{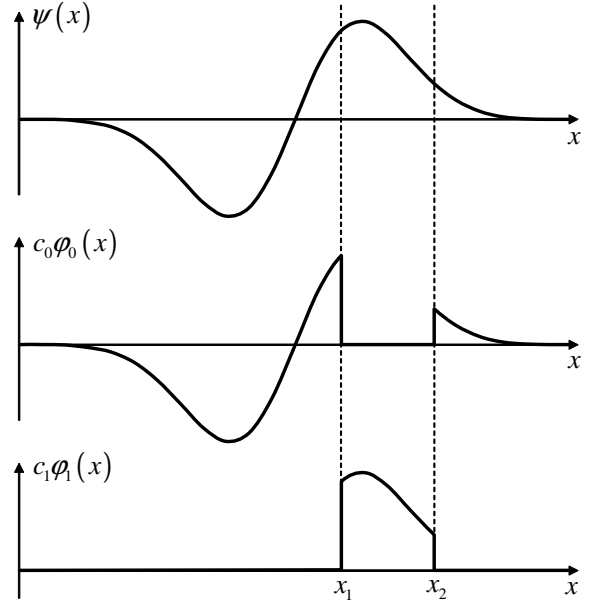}
\caption{Decomposition of the wavefunction $\psi(x)$ into
terms $c_0\varphi_0(x)$ and $c_1\varphi_1(x)$.}
\label{fig:wf}
\end{figure}

It is easy to see that any normalized wavefunction $\psi(x)$ can be represented as
\begin{equation}
\label{pr4-decomposition}
  \psi(x) = c_0 \varphi_0(x) + c_1 \varphi_1(x) ,
\end{equation}
where $c_0$ and $c_1$ are some numbers, $|c_0|^2+|c_1|^2=1$, 
and $\varphi_0(x)$ and $\varphi_1(x)$ are such normalized wavefunctions that 
\begin{gather}
\label{pr4-phi0-property}
  \forall x\in[x_1,x_2]:     \varphi_0(x)=0, \\ 
\label{pr4-phi1-property}
  \forall x\not\in[x_1,x_2]: \varphi_1(x)=0,
\end{gather}
as illustrated in Fig.~\ref{fig:wf}.
Due to the postulated properties of the detector, we have:
\begin{equation}
  \mathcal{P}[\varphi_0] = 0 \quad\text{and}\quad \mathcal{P}[\varphi_1] = 1.
\end{equation}
So we can use Eq.~(\ref{pr4-pclick-answer}), and get 
\begin{equation}
\label{pr4-p}
  \mathcal{P}[\psi] = |c_1|^2. 
\end{equation}

Now let us find $c_1$. First, we represent $\psi(x)$ as the sum $\Phi_0(x)+\Phi_1(x)$, 
where
\begin{align}
  \Phi_0(x) &= \bigl(1-m(x)\bigr) \, \psi(x), \\ 
  \Phi_1(x) &= m(x) \, \psi(x),
\end{align}
and
\begin{equation}
  m(x) = 
  \begin{cases}
    1, \; \text{if} \;\, x\in[x_1,x_2], \\
    0, \;  \text{otherwise}.
  \end{cases}
\end{equation}
Then, we obtain $\varphi_0(x)$ and $\varphi_1(x)$ by normalization of the functions 
$\Phi_0(x)$ and $\Phi_1(x)$:
\begin{equation}
\label{pr4-phi}
  \varphi_0(x) = \Phi_0(x) / c_0, \quad 
  \varphi_1(x) = \Phi_1(x) / c_1, 
\end{equation}
where $c_0$ and $c_1$ are norms of the functions $\Phi_0(x)$ and $\Phi_1(x)$:
\begin{multline}
\label{pr4-c0}
  c_0 = \left( \int_{-\infty}^{+\infty}|\Phi_0(x)|^2dx \right)^{1/2} = \\
  = \left( \int_{-\infty}^{x_1}|\psi(x)|^2dx + \int_{x_2}^{+\infty}|\psi(x)|^2dx \right)^{1/2} ,
\end{multline}
\begin{equation}
\label{pr4-c1}
  c_1 = \left( \int_{-\infty}^{+\infty}|\Phi_1(x)|^2dx \right)^{1/2} = 
  \left( \int_{x_1}^{x_2}|\psi(x)|^2dx \right)^{1/2} .
\end{equation}
It is easy to check that the functions~(\ref{pr4-phi}) obey the properties~(\ref{pr4-phi0-property}), 
(\ref{pr4-phi1-property}), and that the equality~(\ref{pr4-decomposition}) is satisfied.

Finally, substitution of Eq.~(\ref{pr4-c1}) into Eq.~(\ref{pr4-p}) 
gives the following answer for the probability 
of finding a particle in the range $[x_1,x_2]$ of $x$-coordinate:
\begin{equation}
\label{pr4-answer}
  \mathcal{P}[\psi] = \int_{x_1}^{x_2}|\psi(x)|^2dx .
\end{equation}
This is the Born rule in its integral form. Thus, 
the five Assumptions of 
Subsection~\ref{sub:st3-assump} provide enough background for derivation the Born rule.

\section{Necessity of each Assumption} 
\label{sec:di}

It is shown above that the set of five Assumptions 
given in Subsection~\ref{sub:st3-assump} is \emph{sufficient} 
for the derivation of the Born rule. The next question is: are 
these Assumptions \emph{necessary} for the derivation of the Born rule? 

We will prove by contradiction that each Assumption is indeed necessary. 
Namely, for each Assumption we will find a 
\emph{counterexample}---an expression for the probability that differs 
from the Born rule and, at the same time, agrees with all the Assumptions 
\emph{except} the given one. This will prove the necessity of the given Assumption. 

Let $\mathcal{P}_\uparrow^{(0)}(\mathcal{S},\mathcal{A})$ and 
$\mathcal{P}_\downarrow^{(0)}(\mathcal{S},\mathcal{A})$ be ``true'' probabilities 
(obeying the Born rule) of 
the ``spin-up'' and ``spin-down'' outcomes when the spin in the state $\mathcal{S}$ 
is measured by the apparatus $\mathcal{A}$. 
Then, we introduce the quantities
\begin{align}
\label{di-p1-up}
  \mathcal{P}_\uparrow^{(1)}(\mathcal{S},\mathcal{A},x) &= 
  \begin{cases}
    1, \; \text{if} \;\, \mathcal{P}_\uparrow^{(0)}(\mathcal{S},\mathcal{A}) > x, \\
    0, \;  \text{otherwise},
  \end{cases} 
  \\
\label{di-p1-down}
  \mathcal{P}_\downarrow^{(1)}(\mathcal{S},\mathcal{A},x) &= 
  1 - \mathcal{P}_\uparrow^{(1)}(\mathcal{S},\mathcal{A},x),
\end{align}
where $x$ is a random number associated with the given measurement event and uniformly 
distributed in the range $[0,1]$. Let us consider the quantities $\mathcal{P}_\uparrow^{(1)}$ 
and $\mathcal{P}_\downarrow^{(1)}$ as probabilities of ``up'' and ``down'' outcomes. 
One can easily see that they agree with all Assumptions except the first one. At the same time, 
they do not obey the Born rule. So, the expressions~(\ref{di-p1-up}), (\ref{di-p1-down}) 
provide a counterexample proving that Assumption~1 is necessary.

Now we consider a case when Assumption 2 is violated. Let the evolution of some collection of 
$N$ spins is described not by unitary (conserving the scalar product) operators, but by 
operators conserving a ``modified scalar product'' 
\begin{equation}
  \langle\Psi\|\Phi\rangle = \langle \Psi | 
  \underbrace{\hat{A}\otimes\hat{A}\otimes\cdots\otimes\hat{A}}_{N\;\text{times}} 
  | \Psi \rangle
\end{equation}
of any $N$-spin state vectors $\ket{\Psi}$ and $\ket{\Phi}$, 
where $\hat{A}$ is some positive self-adjoint operator in the spin state space. 
If Assumptions~1,3,4,5 remain in force, then, instead of Theorem~1, one can deduce that 
the probability of any outcome is a linear function of the ``modified spin polarization'' 
\begin{equation}
  \frac
  { \langle \mathbf{\Psi} \| \hat{\boldsymbol{\sigma}} 
    \otimes \mathbb{I}_\mathcal{E} | \mathbf{\Psi} \rangle } 
  { \langle \mathbf{\Psi} \| \mathbf{\Psi} \rangle } \,.
\end{equation}
Then, instead of the Born rule~(\ref{pr3-bornrule-pure}), one can get the following probabilities:
\begin{equation}
\label{di-p2}
  \mathcal{P}_\uparrow^{(2)} \bigl( \ket{\psi} \bigr) = 
  \frac{\left| \langle\varphi_\uparrow|\hat{A}|\psi\rangle \right|^2}{\langle\psi|\hat{A}|\psi\rangle} \,, \quad 
  \mathcal{P}_\downarrow^{(2)} \bigl( \ket{\psi} \bigr) = 
  \frac{\left| \langle\varphi_\downarrow|\hat{A}|\psi\rangle \right|^2}{\langle\psi|\hat{A}|\psi\rangle} \,,
\end{equation}
where vectors $\ket{\varphi_\uparrow}$ and $\ket{\varphi_\downarrow}$ obey the relations
\begin{gather*}
  \langle\varphi_\uparrow|\hat{A}|\varphi_\uparrow\rangle = 
  \langle\varphi_\downarrow|\hat{A}|\varphi_\downarrow\rangle = 1, \\
  \langle\varphi_\uparrow|\hat{A}|\varphi_\downarrow\rangle = 0.
\end{gather*}
Hence, necessity of Assumption~2 is proven by a counterexample provided by Eq.~(\ref{di-p2}).

The next expressions can serve as a counterexample proving the necessity of the remaining three Assumptions:
\begin{gather}
\label{di-p3-up}
  \mathcal{P}_\uparrow^{(3)}(\mathcal{S},\mathcal{A}) = 
  \left(3 - 2\, \mathcal{P}_\uparrow^{(0)}(\mathcal{S},\mathcal{A}) \right) 
  \left( \mathcal{P}_\uparrow^{(0)}(\mathcal{S},\mathcal{A}) \right)^2 ,
  \\
\label{di-p3-down}
  \mathcal{P}_\downarrow^{(3)}(\mathcal{S},\mathcal{A}) = 
  \left(3 - 2\, \mathcal{P}_\downarrow^{(0)}(\mathcal{S},\mathcal{A}) \right) 
  \left( \mathcal{P}_\downarrow^{(0)}(\mathcal{S},\mathcal{A}) \right)^2 .
\end{gather}
The quantities $\mathcal{P}_\uparrow^{(3)}$ and $\mathcal{P}_\downarrow^{(3)}$, considered as probabilities 
of ``up'' and ``down'' measurement results, obey Eqs.~(\ref{a1-1}),(\ref{a3-1}),(\ref{a3-3}),(\ref{a4-1}), 
but violate Eq.~(\ref{a5-3}). This fact 
can be interpreted in three different ways. First, if we accept Assumptions~1,2,4,5, then violation of Eq.~(\ref{a5-3}) 
means violation of either the multiplication rule~(\ref{a3-2}), or the addition rule~(\ref{a4-3}); both rules are contained in 
Assumption~3. The second option is to keep Assumptions~1,2,3,5, and attribute the violation of Eq.~(\ref{a5-3}) to 
violation of Eq.~(\ref{a4-2}), which means rejection of the Assumption~4. The third possibility is to accept 
Assumptions~1,2,3,4, and reject Assumption~5; in this case Eq.~(\ref{a5-3}) is no more relevant, therefore the 
probabilities $\mathcal{P}_\uparrow^{(3)}$ and $\mathcal{P}_\downarrow^{(3)}$ are in accordance with 
the first four Assumptions.

Thus, if we sacrifice Assumption~3, or~4, or~5, then the probabilities~(\ref{di-p3-up}), (\ref{di-p3-down}) 
would satisfy the remaining Assumptions but disargee with the Born rule. So Eqs.~(\ref{di-p3-up}) 
and (\ref{di-p3-down}) provide a counterexample demonstrating necessity of assumptions~3,4,5.

Now it is proven that each of the five Assumpltions is indeed necessary for the derivation of the Born rule.

\section{Conclusions} \label{sec:co}

In this paper, the Born rule is deduced from five very simple and reasonable assumptions 
given in Subsection~\ref{sub:st3-assump}. Each of these assumptions is necessary, as 
shown in Section~\ref{sec:di}. Unlike previous studies, our approach treats measuring 
devices as black boxes (no analysis of their operation is required), and allows to 
consider non-ideal devices (which are not described by operators). This advantage 
makes it possible to get more than the Born rule---namely, we show that 
probabilities of outcomes of any (ideal or non-ideal) measurement on the spin 1/2 
are linear functions of the spin polarization vector (Theorem~1). This result is 
enough for associating a positive-operator valued measure (POVM) to any measurement 
on the spin 1/2, see Eq.~(\ref{pr2-povm}). 
The Born rule comes in the particular case of ideal measurements, 
considered in Theorem~2. Though we are concentrated on spin-1/2 measurements, our results 
can be easily generalized, as illustrated in Subsection~\ref{sub:pr4-coord} for the 
case of the position measurement.

\bibliography{envariance}

\end{document}